\documentclass{amsart}
\usepackage{graphicx,amssymb}
\newtheorem{thm}{Theorem}[section]
\newtheorem{cor}[thm]{Corollary}

\theoremstyle{definition}

\theoremstyle{remark}

\numberwithin{equation}{section}

\newcommand{\av}[1]{\left\langle{#1}\right\rangle}
\newcommand{\mbold}[1]{\mbox{\boldmath${#1}$}}
\newcommand{\re}{\mathrm{Re}}
\renewcommand{\d}{\mathrm{d}}
\def\beq{\begin{eqnarray}}
\def\eeq{\end{eqnarray}}
\def\beqs{\begin{equation}\begin{split}}
\def\eeqs{\end{split}\end{equation}}
\begin{document}

\title{Late-time behaviour of the tilted Bianchi type VI$_h$ models}
 \author[S Hervik, R J van den Hoogen, W C Lim and A A Coley]{S Hervik$^{1}$, R J van den Hoogen$^{2, 1}$, W C Lim$^{1, 3, 4}$ and A A Coley$^{1}$ }%
\address{$^{1}$ Department of Mathematics \& Statistics, Dalhousie University,
Halifax, Nova Scotia, Canada B3H 3J5}%
\address{$^2$ Department of Mathematics, Statistics and Computer Science,
St. Francis Xavier University, Antigonish, Nova Scotia, Canada B2G 2W5}
\address{$^3$ Department of Physics, University of Alberta, Edmonton, Alberta, Canada T6G 2G7}
\address{$^4$ Department of Physics, Princeton University, Princeton, New Jersey, USA 08544}
\email{herviks@mathstat.dal.ca, rvandenh@stfx.ca, wlim@princeton.edu, aac@mathstat.dal.ca}%

\subjclass{}%
\keywords{}%

\date{\today}%
\begin{abstract}

We study tilted perfect fluid cosmological models with a constant
equation of state parameter in spatially homogeneous models of
Bianchi type VI$_h$ using dynamical systems methods and numerical
experimentation, with an emphasis on their future asymptotic
evolution. We determine all of the equilibrium points of the type
VI$_h$ state space (which correspond to exact self-similar
solutions of the Einstein equations, some of which are new), and their stability is investigated. We find that
there are vacuum plane-wave solutions that act as future
attractors. In the parameter space, a `loophole' is shown to exist in which there are no
stable equilibrium points. We then show that a Hopf-bifurcation
can occur resulting in a stable closed orbit (which we refer to as
the Mussel attractor) corresponding to points both inside the
loophole and points just outside the loophole; in the former case
the closed curves act as late-time attractors while in the latter
case these attracting curves will co-exist with attracting
equilibrium points. In the  special Bianchi type III case, centre
manifold theory is required to determine the future attractors.
Comprehensive numerical experiments are carried out to complement
and confirm the analytical results presented. We note that the
Bianchi type VI$_h$ case is of particular interest in that it
contains many different subcases which exhibit many of the
different possible future asymptotic behaviours of Bianchi
cosmological models.

\end{abstract}

\maketitle

\section{Introduction}

In this paper, we shall study tilted perfect fluid cosmological models with a constant
equation of state parameter, $\gamma$, in spatially homogeneous Bianchi models 
of type VI$_h$ using the formalism
introduced in  \cite{CH2}.  We introduce expansion-normalised
variables \cite{DS1}, determine  the  equilibrium points and their stability properties and consequently investigate the future asymptotic behaviour of the models and
determine the late-time asymptotic states, using  
dynamical systems methods and numerical experimentation. 
In spatially homogeneous cosmological models the
universe is foliated into space-like hypersurfaces (defined by the
group orbits of the respective model) \cite{DS1,EM,DS2,BS,BN}.  For
these perfect fluid models there are two naturally defined time-like vector
fields (i.e., congruences): the unit vector field, $n^{\mu}$, normal
to the group orbits and hence orthogonal to the surfaces of
transitivity (the 'geometric' congruence), and the four-velocity,
$u^{\mu}$, of the fluid (the 'matter' congruence). If $u^{\mu}$ is not aligned with $n^{\mu}$, the model is
called \emph{tilted} (and non-tilted or orthogonal otherwise)
\cite{KingEllis}. The geometric congruence is necessarily
geodesic, vorticity-free and acceleration-free. The matter
congruence, on the other hand, is not necessarily geodesic and
can have both vorticity and acceleration. 

We shall follow convention and use the kinematical
quantities associated with the normal congruence $n^\mu$ 
(rather than the fluid flow $u^\mu$) of the
spatial symmetry surfaces as variables. This avoids the possible singular behaviour experienced by the fluid observers in these models \cite{tilt1,tilt2,tilt3,tilt4} \footnote{In these papers the physical properties of these models, and particularly the observations and singularity structure in models in
which the tilt  becomes extreme asymptotically (i.e., $v^2 \rightarrow 1$) as
measured by observers moving with the geometric congruence, was studied.} (these papers also give the appropriate boost-formluae relating the kinematic variables and curvatures in the two frames). In principle, it is also necessary to investigate the behaviours in the fluid frame as the asymptotes may differ (which is the case for so-called 'whimper singularities' \cite{EllisKing}); however, this will not be pursued here (criteria for when this may happen are given in \cite{tilt4}). We will also focus on the dynamical behaviour rather than the physical effects of these models; observational consequences of (tilted) Bianchi models have been studied in, for example, \cite{Jaffe1,Jaffe2,ColeyLim}.

 We
will assume a perfect-fluid matter source with $p=(\gamma-1)\mu$ as equation of
state, where $\mu$ is the energy density, $p$ is the pressure, and $\gamma$ is a
constant.  Causality then requires $\gamma$ to be in the interval
$0\leq\gamma\leq2$.  A positive cosmological constant may also be included in the
models.  Tilted SH cosmologies with a $\gamma$-law perfect fluid source have
been studied by a number of authors. In \cite{BHtilted} the stability of non-tilted universes with respect to tilt was studied and Table \ref{PrevStudies} shows previous studies of tilted universes of Bianchi type II-VIII. 
It has been proven that if the matter obeys the strong energy condition, the
positive pressure criteria, the dominant energy condition, and a matter
regularity condition all
Bianchi type IX models  recollapse to the future \cite{BGT,BT,LinWald1,LinWald2}. 
\begin{table}
\caption{Previous studies of the dynamical behaviour of tilted Bianchi models}\label{PrevStudies}
\begin{tabular}{|c|c|c|}
\hline
Bianchi type & Type of tilt & References \\
\hline \hline
II & General & \cite{HBWII} \\
\hline
IV & General & \cite{CH2,HHC} \\
\hline
V  & Irrotational & \cite{Shikin,Collins,CollinsEllis,HWV} \\
 & General & \cite{Harnett,CH2}\\
\hline 
VI$_0$ & General & \cite{hervik}\\
 & 2 fluids & \cite{CH1} \\
\hline
VI$_h$ & Subset & \cite{CH2} \\
\hline 
VII$_h$ & General & \cite{HHC}\\
\hline 
VII$_0$ & Irrotational & \cite{CH2,LimDW}\\
    & General & \cite{HHLC} \\
\hline 
VIII & General & \cite{HLim}\\
\hline
\end{tabular}
\end{table}

Of the most general ever-expanding Bianchi models (namely VI$_h$, VII$_h$ and VIII) only the dynamics of the type VI$_h$ models remain to be studied. This paper aims to fill this gap in the study of the behaviour of general spatially homogeneous Bianchi models.  The behaviour of these models has been shown to be both interesting and surprising. In particular, for the Bianchi type VIII models (and Bianchi models of type VII$_0$
\cite{HHLC}), the state space is unbounded and consequently, for all
non-inflationary perfect fluids, one of the curvature variables grows without
bound at late times.  It was found that in Bianchi type VIII models
\cite{HLim} with fluids stiffer than dust
($1<\gamma<2$), the fluid will in general tend towards a state of extreme tilt,
while for fluids less stiff than dust ($0<\gamma \leq 1$) the fluid will in the
future be asymptotically non-tilted.  Using both dynamical systems theory and a
detailed numerical analysis, the late-time behaviour of tilting perfect fluid
Bianchi models of types IV and VII$_h$ was studied, and it was found that the
plane waves are the only future attracting equilibrium points for
non-inflationary fluids in Bianchi type VII$_h$ models \cite{HHC}.  A tiny region of
parameter space (the ``loophole'') in the Bianchi type IV model was shown to contain
a closed orbit which acts as an attractor \cite{CH2}.  From a detailed numerical
analysis it was found that at late times the normalised energy-density tends to
zero and the normalised variables 'freeze' into their asymptotic values, and it
was then shown that there is an open set of parameter space in the type VII$_h$
models in which solution curves approach a compact surface that is topologically
a torus \cite{HHC}.

In this work we determine all of the equilibrium points of the type VI$_h$ state space. These equilibrium points correspond to exact self-similar solutions of the Einstein equations and play a special role in the general evolution of the system \cite{carrcoley}. In particular, the stability of these solutions are determined. Some of these solutions are new (and one is given implicitly). A complete catalogue of self-similar solutions is given and the late time attractors are classified. A detailed numerical analysis is included, complementing the analytical results. We note that the Bianchi type VI$_h$ case is very complicated, with many cases to consider, and the resulting dynamics include many of the different types of behaviour described in previous work.  

This paper is organised as follows. In the following section we present the equations of motion for the tilted Bianchi type VI$_h$ models and discuss the state space and various invariant sets of physical interest. In section  \ref{sect:Qual} we present monotonic functions and give a list of all equilibrium points. Section \ref{sect:late-time} is devoted to the analysis of the late-time behaviour and in section \ref{sect:num} we give a synopsis of the numerical analysis done. The final section is devoted to a discussion of the ramifications of our results.


\section{Equations of motion}
\subsection{The orthonormal frame approach}
The line-element of a Bianchi cosmology can be written
\beq
\d s^2=-\d t^2+\delta_{ab}{\mbold\omega}^a{\mbold\omega}^b,
\eeq 
where $t$ is the co-moving cosmological time. The one-forms ${\mbold\omega}^a$ are left-invariant one-forms on the hypersurfaces spanned by the group orbits. They obey the commutator relation
\beq
\left(\d{\mbold\omega}^a\right)_{\perp}=-\frac 12C^a_{~bc}{\mbold\omega}^b\wedge{\mbold\omega}^c, \quad C^a_{~bc}=\epsilon_{bcd}n^{da}+a_{b}\delta^a_{~c}-a_{c}\delta^a_{~b},
\eeq
where $n^{ab}$ is a symmetric tensor, $a_c=-(1/2)C^a_{~ac}$ and $\perp$ means projection onto the spatial hypersurfaces. Furthermore, the Jacobi identity implies $n^{bc}a_c=0$. 

Since $n^{ab}$ is a symmetric spatial tensor we can characterise it in terms of its eigenvalues, $(n_1,n_2,n_3)$. We are interested in the Bianchi type VI$_h$ models for which $a_c\neq 0$, and hence, one of the eigenvalues of $n^{ab}$ is necessarily zero: $n_1=0$ (say). Moreover, for the Bianchi type VI$_h$ models we have the relation $a_ca^c=hn_2n_3$, which defines the group parameter $h$. 

The geometric (or normal) congruence, $n^{\mu}$, is given by ${\bf n}=\partial/\partial t$. It is also useful to define the shear and the Hubble scalar associated with the congruence $n^\mu$: 
\beq
H\equiv\frac 13n^\mu_{~;\mu}, \quad \sigma_{\mu\nu}\equiv n_{\mu;\nu}-H h_{\mu\nu}, 
\eeq
where $h_{\mu\nu}$ is the spatial metric on the hypersurfaces spanned by the group orbits. 
The matter variables are chosen to be the energy density, $\mu$, and the tilt-velocity, $v^a$, which is defined as the 3-velocity of the fluid with respect to the geometric (or normal) congruence, $n^{\mu}$. 
The equations of motion can now be written down in terms of the Hubble scalar, $H;$ the shear, $\sigma_{ab}$; the curvature variables $n^{ab}$ and $a_c$; and the matter variables $\mu$ and $v^a$.  

In the dynamical systems approach it is common to introduce expansion-normalised variables (we divide the variables with the appropriate powers of $H$).   
The paper \cite{CH2} contains all the details regarding the  determination of the evolution 
equations for the tilted cosmological models under consideration. 
These equations, written in {\em gauge invariant} form, allow one to choose 
the gauge that is best suited to the application at hand. Here, we shall adopt the $N$-gauge in which the function 
${\bf N}_{\times}$ is purely imaginary; this is ensured by the choice
$\phi'=\sqrt{3}\lambda\Sigma_-$, where $\lambda$ is defined by
$\bar{N}=\lambda\mathrm{Im}({\bf N}_{\times})$. The evolution equation for
$\bar{N}$ can then be replaced with an evolution equation for $\lambda$,
which ensures a closed system of equations.  For a qualitative analysis of these models, the $N$-gauge
is preferable since the resulting dynamical system is well defined in the Bianchi VI$_h$ case, in particular, $|\lambda|<1$. The relation between these variables (and their corresponding differential equations) and the models they represent are discussed in more detail in \cite{DS1}. 

In the notation of \cite{CH2}, the expansion-normalised anisotropy and curvature variables used in this paper are:
\[ {\mbold\Sigma}_1=\Sigma_{12}+i\Sigma_{13}, \quad {\mbold\Sigma}_{\times}=\Sigma_-+i\Sigma_{23}, \quad {\bf N}_{\times}=iN. \]
We will also adopt the dimensionless time parameter $\tau$, which is related to the cosmological time $t$ via $\mathrm{d} t/\mathrm{d}\tau =(1/H)$, where $H$ is the Hubble scalar.

Using expansion-normalised variables, the equations of motion in the $N$-gauge
are (see \cite{CH2} for the complete derivation of the equations): 
\beq 
\Sigma_+'&=& (q-2)\Sigma_++{3}(\Sigma_{12}^2+\Sigma^2_{13})-2N^2 +\frac{\gamma\Omega}{2G_+}\left(-2v_1^2+v_2^2+v_3^2\right) \\
\Sigma_-'&=&(q-2-2\sqrt{3}\Sigma_{23}\lambda)\Sigma_-+\sqrt{3}(\Sigma_{12}^2-\Sigma_{13}^2) +2AN +\frac{\sqrt{3}\gamma\Omega}{2G_+}\left(v_2^2-v_3^2\right)\\
\Sigma'_{12}&=& \left(q-2-3\Sigma_+-\sqrt{3}\Sigma_-\right)\Sigma_{12} -\sqrt{3}\left(\Sigma_{23}+\Sigma_-\lambda\right)\Sigma_{13} +\frac{\sqrt{3}\gamma\Omega}{G_+}v_1v_2\\
\Sigma'_{13}&=&\left(q-2-3\Sigma_++\sqrt{3}\Sigma_-\right)\Sigma_{13}-\sqrt{3}\left(\Sigma_{23}-\Sigma_-\lambda\right)\Sigma_{12}+\frac{\sqrt{3}\gamma\Omega}{G_+}v_1v_3\\
\Sigma'_{23}&=&(q-2)\Sigma_{23}-2\sqrt{3}N^2\lambda+2\sqrt{3}\lambda\Sigma_-^2+2\sqrt{3}\Sigma_{12}\Sigma_{13}+ \frac{\sqrt{3}\gamma\Omega}{G_+}v_2v_3\\
N'&=& \left(q+2\Sigma_++2\sqrt{3}\Sigma_{23}\lambda\right){N}\\
\lambda' &=& 2\sqrt{3}\Sigma_{23}\left(1-\lambda^2\right)\\
A'&=& (q+2\Sigma_+)A . 
\eeq 
The equations for the fluids are:
\beq
\Omega'&=& \frac{\Omega}{G_+}\Big\{2q-(3\gamma-2)+2\gamma Av_1
 +\left[2q(\gamma-1)-(2-\gamma)-\gamma\mathcal{S}\right]V^2\Big\}
 \quad \\
 v_1' &=& \left(T+2\Sigma_+\right)v_1-2\sqrt{3}\Sigma_{13}v_3-2\sqrt{3}\Sigma_{12}v_2-A\left(v_2^2+v_3^2\right)-\sqrt{3}N\left(v_2^2-v_3^2\right)\\
 v_2'&=& \left(T-\Sigma_+-\sqrt{3}\Sigma_-\right)v_2-\sqrt{3}\left(\Sigma_{23}+\Sigma_-\lambda\right)v_3+\sqrt{3}\lambda{N}v_1v_3+\left(A+\sqrt{3}N\right)v_1v_2 \\
 v_3'&=& \left(T-\Sigma_++\sqrt{3}\Sigma_-\right)v_3-\sqrt{3}\left(\Sigma_{23}-\Sigma_-\lambda\right)v_2-\sqrt{3}\lambda{N}v_1v_2+\left(A-\sqrt{3}N\right)v_1v_3 \\
 V'&=&\frac{V(1-V^2)}{1-(\gamma-1)V^2}\left[(3\gamma-4)-2(\gamma-1)Av_1-\mathcal{S}\right],
\eeq 
where 
\beq q&=& 2\Sigma^2+\frac
12\frac{(3\gamma-2)+(2-\gamma)V^2}{1+(\gamma-1)V^2}\Omega\nonumber \\
\Sigma^2 &=& \Sigma_+^2+\Sigma_-^2+\Sigma_{12}^2+ \Sigma_{13}^2+\Sigma_{23}^2\nonumber \\
\mathcal{S} &=& \Sigma_{ab}c^ac^b, \quad c^ac_{a}=1, \quad v^a=Vc^a,\quad \nonumber \\
 V^2 &=& v_1^2+v_2^2+v_3^2,\quad  \nonumber \\
 T&=& \frac{\left[(3\gamma-4)-2(\gamma-1)Av_1\right](1-V^2)+(2-\gamma)V^2\mathcal{S}}{1-(\gamma-1)V^2}\nonumber\\
 G_+&=&1+(\gamma-1)V^2\nonumber.
\eeq 
These variables are subject to the constraints 
\beq
1&=& \Sigma^2+A^2+N^2+\Omega \label{const:H}\\
0 &=& 2\Sigma_+A+2\Sigma_-N+\frac{\gamma\Omega v_1}{G_+} \label{const:v1}\\
0 &=&
-\left[\Sigma_{12}(N+\sqrt{3}A)+\Sigma_{13}\lambda{N}\right]+\frac{\gamma\Omega v_2}{G_+} \label{const:v2}\\
0 &=&
\left[\Sigma_{13}(N-\sqrt{3}A)+\Sigma_{12}\lambda{N}\right]+\frac{\gamma\Omega v_3}{G_+} \label{const:v3} \\
0&=& A^2+3h\left(1-\lambda^2\right)N^2.\label{const:group} 
\eeq 
The parameter $\gamma$ will be assumed to be in the interval $\gamma\in (0,2)$. 
The generalized Friedmann equation (\ref{const:H}) yields an expression which effectively defines the energy density $\Omega$. We will assume that this energy density is non-negative: $\Omega\geq 0$. 
Therefore, the state vector can thus be considered 
${\sf X}=[\Sigma_+,\Sigma_-,\Sigma_{12},\Sigma_{13},\Sigma_{23},N,\lambda,A,v_1,v_2,v_3]$ 
modulo the constraint equations  (\ref{const:v1})-(\ref{const:group}). 
Thus the dimension of the physical state space is seven (for a given value of the parameter $h$).  
Additional details are presented in \cite{CH2}.

The dynamical system is invariant under the following discrete symmetries :
$$\begin{tabular}{l}
$\phi_1:~[\Sigma_+,\Sigma_-,\Sigma_{12},\Sigma_{13},\Sigma_{23},N,\lambda,A,v_1,v_2,v_3] 
\mapsto  [\Sigma_+,\Sigma_-,\Sigma_{12},\Sigma_{13},\Sigma_{23},-N,\lambda,-A,-v_1,-v_2,-v_3] $ \\
$\phi_2:~[\Sigma_+,\Sigma_-,\Sigma_{12},\Sigma_{13},\Sigma_{23},N,\lambda,A,v_1,v_2,v_3] 
\mapsto [\Sigma_+,-\Sigma_-,\Sigma_{13},\Sigma_{12},\Sigma_{23},-N,\lambda,A,v_1,v_3,v_2] $ \\
$\phi_3^{\pm}\! :~[\Sigma_+,\Sigma_-,\Sigma_{12},\Sigma_{13},\Sigma_{23},N,\lambda,A,v_1,v_2,v_3]
\mapsto [\Sigma_+,\Sigma_-,\pm \Sigma_{12},\mp \Sigma_{13},-\Sigma_{23},N,-\lambda,A,v_1,\pm v_2,\mp v_3]$\\
$\phi_4:~[\Sigma_+,\Sigma_-,\Sigma_{12},\Sigma_{13},\Sigma_{23},N,\lambda,A,v_1,v_2,v_3]
\mapsto[\Sigma_+,\Sigma_-,-\Sigma_{12},-\Sigma_{13},\Sigma_{23},N,\lambda,A,v_1,-v_2,-v_3]$
\end{tabular}$$
These discrete symmetries imply that without loss of  generality we can restrict the variables $A\geq0$, and $N\geq0$,  since the dynamics in the other regions can be obtained by simply applying one or more of  the maps above.  The third and fourth symmetries listed imply that one can add  additional constraints on the variables $\Sigma_{12},\Sigma_{13},v_2$ or $v_3$; however, in general there is no natural way to restrict any one of the variables, and hence we will not do so here.  Note that any equilibrium point in the region $v_2>0$ has a matching equilibrium point in the region $v_2<0$. 

\subsection{Invariant sets}
In this analysis we will be concerned with the following invariant sets: 
\begin{enumerate}
\item{} $T(VI_h)$: The general tilted type VI$_h$  model: $ |\lambda|<1$. 
\item{} $T_1(VI_h)$: A one-tilted type VI$_h$ model: $|\lambda|<1$, $v_2=v_3=\Sigma_{12}=\Sigma_{13}=0$. 
\item{} $T_{1,0}(VI_h)$: A one-tilted diagonal type VI$_h$ model: $v_2=v_3=\Sigma_{12}=\Sigma_{13}=\Sigma_{23}=\lambda=0$.
\item{} $N^{\pm}(VI_h)$: A class of tilted type VI$_h$ models with $W^0=0$ (see eq.(\ref{vorticity}) for definition):  $|\lambda|<1$, $N_{ab}v^av^b=0$ ($h\neq -1/9)$. 
\item{} $T_2^+(VI_h)$: A two-tilted type VI$_h$ model. This is the fixed-point-set of $\phi_3^+$ and is given by  $v_3=\Sigma_{13}=\Sigma_{23}=\lambda=0$.
\item{} $T_2^-(VI_h)$: A two-tilted type VI$_h$ model. This is the fixed-point-set of $\phi^-_3$ and is given by $v_2=\Sigma_{12}=\Sigma_{23}=\lambda=0$.
\item{} $B(VI_h)$: Non-tilted type VI$_h$: $|\lambda|<1$, $v_1=v_2=v_3=\Sigma_{12}=\Sigma_{13}=0$.
\item{} $B_0(VI_h)$: A class of diagonal non-tilted type VI$_h$ models ($n^{\alpha}_{~\alpha}=0$): $V=\Sigma_{12}=\Sigma_{13}=\Sigma_{23}=\lambda=0$
\item{} $T(II)$: The general type II model: $\lambda=\pm 1$, $A=0$. 
\item{}$B(I)$: Type I: ${N}=A=V=0$.
\item{}{$\partial T(I)$}: ``Tilted'' vacuum type I: $\Omega=N=A=0$.
\end{enumerate}
Regarding $N^{\pm}(VI_h)$, this invariant set is not a manifold; it is similar to the light-cone in 2-dimensional Minkowski space. Therefore, $N^{\pm}(VI_h)-T_1(VI_h)$ consists of 4 disconnected pieces. By the symmetry $\phi_4$, these are actually only two inequivalent pieces. Here, we choose $N^{\pm}(VI_h)$ such that 
\[ T_2^+(VI_h)\subset N^+(VI_h), \quad T_2^-(VI_h)\subset N^-(VI_h) \]  
Since $N^+(VI_h)\cap N^-(VI_h)=T_1(VI_h)$, both $N^+(VI_h)$ and $N^-(VI_h)$ are invariant sets. 

We note that the closure of the set $T(VI_h)$ is given by
\beq
\overline{T(VI_h)}&=&T(VI_h)\cup T(II)\cup B(I) \cup \partial T(I).
\label{eq:decomp}\eeq
Since the boundaries may play an important role in the evolution of the dynamical system  we must consider all of the sets in the decomposition (\ref{eq:decomp}).

\subsection{The case $h=-1/9$} A few comments  are in order when $h=-1/9$. Let us consider the constraint equations (\ref{const:v2}) and (\ref{const:v3}) as a linear map
\[ {\sf L}:~(\Sigma_{12},\Sigma_{13})\mapsto (v_2,v_3)/G_+,\]
where ${\sf L}$ is considered given in terms of $A$, $N$, $\lambda$ and $\Omega$. For $h\neq -1/9$, $\det({\sf L})\neq 0$ and the image of ${\sf L}$ is 2-dimensional. However, for $h= -1/9$, $\det({\sf L})= 0$ and the image of ${\sf L}$ is 1-dimensional; hence, in this sense \emph{the constraint equations are degenerate}. This implies that $(v_2,v_3)$ has to be restricted to a 1-dimensional submanifold. We will therefore say that the general type VI$_{-1/9}$ model only allows for 2 tilt degrees of freedom. 

On the same token, this also mean that $(v_2,v_3)=0$ does not necessarily imply that $(\Sigma_{12},\Sigma_{13})$ is zero. In particular, for the non-tilted models this implies that we may have an additional shear degree of freedom; these models have usually been called the exceptional case and are denoted with an asterisk: $B(VI^*_{-1/9})$. 

It is advantageous to redefine $N^{\pm}(VI_h)$ for $h=-1/9$ because, as can be shown, $N_{ab}v^av^b$ is identically zero for $h=-1/9$. Let us instead define 
\[ \widehat{D}=\left[\lambda(\Sigma_{12}^2+\Sigma_{13}^2)+2\Sigma_{12}\Sigma_{13}\right], \] 
and define $N^{\pm}(VI_{-1/9})$ as the set of points where $\widehat{D}=0$. This is the natural generalisation of $N^{\pm}(VI_h)$ to $h=-1/9$.
We also note that for the tilted models, there is an exceptional case of the one-tilted models $T_1(VI_{-1/9})$ which could be denoted $T_1(VI^*_{-1/9})$. However, $T_1(VI^*_{-1/9})=N^-(VI_{-1/9})$ where $N^-(VI_{-1/9})$ is defined above. Therefore, we keep the notation $N^-(VI_{-1/9})$.

\begin{table}
\caption{The dimensions of the invariant sets for $h\neq -1/9$. The group parameter $h$ is regarded as fixed. }
\begin{tabular}{|c||c|c|c|c|c|c|}
\hline
Dim &2 & 3 & 4&5& 6& 7 \\
\hline 
 & $B_0(VI_h)$ &$T_{1,0}(VI_h)$ & $B(VI_h)$ & $T_1(VI_h)$ & $N^{\pm}(VI_h)$ & $T(VI_h)$ \\
& & & $T_2^{\pm}(VI_h)$ & &  &  \\
\hline 
\end{tabular}\\
\end{table}
\begin{table}
\caption{The dimensions of the invariant sets for $h=-1/9$ in terms of the number of tilt degrees of freedom. The left-most column indicates the specialization in terms of the $G_2$ cosmologies. Here, HO means hypersurface orthogonal and an asterisk indicates the exceptional case where the models aquire an addition shear degree of freedom. }
\begin{tabular}{|c||c|c|c|c|c|c|}
\hline
Dim &2 & 3 & 4&5& 6& 7 \\
\hline 
Gen $G_2$ &  &  & vacuum$^*$ & 0$^*$ tilt  & 1$^*$, 2 tilt & 2 tilt \\
HO KVF & vacuum$^*$ & 0$^*$ tilt & 1$^*$ tilt & & &   \\
\hline 
\end{tabular}\\
\end{table}
\subsection{Fluid Vorticity} 
The various invariant subspaces can also be categorised in terms of
the ($H_{\mathrm{fluid}}$-normalised where $H_{\mathrm{fluid}}\equiv (1/3)u^{\mu}_{~;\mu}$) fluid vorticity, $W^{\alpha}$. The vorticity of the fluid for the type VI$_h$ models is given by:
\beq
W_a=\frac{1}{2B}\left(N_{ab}v^b+\varepsilon_{abc}v^bA^c+\frac 1{1-V^2}N_{bc}v^bv^cv_a\right), \quad W_0=-v^aW_a, 
\label{vorticity}\eeq
where 
\[ B\equiv\frac{1-\frac 13(V^2+V^2\mathcal{S}+2A_av^a)}{\sqrt{1-V^2}[1-(\gamma-1)V^2]}.\]
For the invariant sets: 
\begin{enumerate}
\item{} $T(VI_h)$: General vortic type VI$_h$ where all components $W^{\alpha}$ can be non-zero.
\item{} $N^{\pm}(VI_h)$: $W^0=W^1=0$. 
\item{} $T_2^+(VI_h)$: $W^0=W^1=W^2=0$.
\item{} $T_2^-(VI_h)$: $W^0=W^1=W^3=0$.
\item{} $T_1(VI_h)$: $W^0=W^a=0$, non-vortic.
\item{} $B(VI_h)$: $W^0=W^a=0$, non-tilted and non-vortic.
\end{enumerate}
In the special case of $h=-1/9$ further components of the vorticity are zero due to the degeneracy of the constraint equations as explained above. 

\section{Qualitative behaviour}
\label{sect:Qual}
\subsection{Monotone functions} 
There are a number of monotone functions in the state space of interest. For $0<\gamma\leq 6/7$, there exists a monotonically increasing function $Z_1$ defined by
\beq
Z_1&\equiv& \alpha\Omega^{1-\gamma}, \quad \alpha=\frac{(1-V^2)^{\frac 12(2-\gamma)}}{G^{1-\gamma}_+V^{\gamma}_{\phantom{+}}}, \\
Z_1'&=&\left[2(1-\gamma)q+(2-\gamma)+{\gamma}\mathcal{S}\right]Z_1.\nonumber 
\eeq
To see that this is a monotonically increasing function, note first that $|\mathcal{S}|\leq 2\Sigma$ \cite{hervik}. Then, using the constraint equation (\ref{const:H}) and $(1-\Sigma)\geq (1-\Sigma^2)/2$, we can write
\beq
&& 2(1-\gamma)q+(2-\gamma)+{\gamma}\mathcal{S}\nonumber\\ 
&&\geq (6-7\gamma)\Sigma^2+2(1-\gamma)(\left|{\bf N}_{\times}\right|^2+A^2)+\frac{\gamma(1-\gamma)(V^2+3)}{G_+}\Omega , 
\eeq
which is strictly positive for $\gamma<6/7$. 
Thus for $0<\gamma\leq 6/7$, $Z_1$ is monotonically increasing as claimed. 
\begin{thm}\label{thm:nontilted}
For $0<\gamma\leq 6/7$, all tilted Bianchi models  (with $\Omega>0$, $V<1$) of solvable type are asymptotically non-tilted at late times. 
\end{thm}
\begin{proof}
Use of the monotonic function $Z_1$. 
\end{proof}
In fact, the result that these models are asymptotically non-tilted at late times is true for all $\gamma<1$; this follows from further analysis and numerics, but is not covered by this theorem. This theorem is, in fact, true for all ever-expanding Bianchi  models, including ever-expanding class A models. 

An immediate result of the above is the following corollary:
\begin{cor}[Cosmic no-hair]
For $\Omega>0$, $V<1$, and $0<\gamma<2/3$ we have that
\[ \lim_{\tau\rightarrow \infty}\Omega=1, \quad \lim_{\tau\rightarrow \infty}V=0.\]
\label{no-hair}
\end{cor}
\begin{proof} See \cite{CH2}. \end{proof}

Let us define 
\beq 
D\equiv \bar{N}|{\bf v}|^2+\mathrm{Re}({\bf N}_{\times}^*{\bf v}^2),
\eeq
for which 
\[ D'=(q+2T+2Av_1)D.\] 
This implies that $D=0$ is an invariant subspace and defines $N^{\pm}(VI_h)$. Moreover, the following function is a monotone function in $T(VI_h)$, $h\neq -1/9$:
\beq
Z_2&=&\frac{G_+D^2}{(1-V^2)^{\frac 52(2-\gamma)}\Omega}, \\
Z_2'&=&(5\gamma-6)(3-2Av_1)Z_2.
\eeq
This function is monotonically decreasing for $\gamma<6/5$ and monotonically increasing for $6/5<\gamma$. 

For a monotone function which is also monotone for $h=-1/9$ we define 
\beq
\widetilde{D}=(1-\lambda^2)\left[\lambda(\Sigma_{12}^2+\Sigma_{13}^2)+2\Sigma_{12}\Sigma_{13}\right]\frac{N^3G_+^2}{\gamma^2\Omega^2}.
\eeq
Using the constraint equations we have $D=-(1+9h)\widetilde{D}$ so using $\widetilde{D}$ instead of $D$ in $Z_2$ gives the same equation of motion. 

In the subspaces $T_1(VI_h)$ and $N^-(VI_{-1/9})$ we have the monotone function: 
\beq
Z_3&=&\frac{v_1^2\Omega}{A^2G_+(1-V^2)^{\frac 12(2-\gamma)}}, \\
Z_3'&=&-(2-\gamma)(3-2Av_1)Z_3.
\eeq
This function is monotonically decreasing in $T_1(VI_h)$. 

The monotonic function $Z_3$ immediately implies:
\begin{thm}[Future behaviour in $T_1(VI_h)$ and $N^-(VI_{-1/9})$]
For $2/3<\gamma<2$, $\Omega>0$, $A>0$, $v_1^2<1$, $v_2=v_3=0$ we have that:  
\[ \text{either}\quad \lim_{\tau\rightarrow \infty}\Omega=0, \quad
\text{or}\quad  \lim_{\tau\rightarrow \infty}V=0.\]
\end{thm}
This implies that all non-vortic type VI$_h$ universes are either asymptotically vacuum or non-tilted at late times. 

In the remains of the section we will present all the equilibrium points including their eigenvalues, and discuss their stability. 
\subsection{Equilibrium points} 
The Bianchi type VI$_h$ models possess a wealth of equilibrium points. A full catalogue of the equilibrium points has up until now been lacking\footnote{The list in Apostolopoulos' paper \cite{Apo2} is incomplete.}. Therefore, several of the equilibrium points we list are new. Due to the function $Z_2$, the type VI$_h$ equilibrium points can be divided into 4 cases, according to whether $D=0$, $\Omega=0$ (vacuum), $\gamma=6/5$, or $V=1$ (extremely tilted).  The equilibrium points in $\partial T(I)$ are all unstable and will not be given here.

\subsubsection{$B(I)$: Equilibrium points of Bianchi type I}  
\begin{enumerate} 
\item{}$\mathcal{I}(I)$: $\Sigma_+=\Sigma_-=\Sigma_{12}=\Sigma_{13}=\Sigma_{23}=A=N=V=0$ and $\Omega=1$. Here, $|\lambda|<1$ and is an unphysical parameter.  This represents the flat Friedman-Lema{\^i}tre model. 
\paragraph{Eigenvalues:} 
\[ -\frac{3(2-\gamma)}{2} [\times 5],~ \frac{3\gamma-2}{2} [\times 2].\] 
\end{enumerate}

\subsubsection{$T(II)$: Equilibrium points of Bianchi type II} 
All of the tilted equilibrium points come in pairs. These represent identical solutions (they differ by a frame rotation); however, since their embeddings in the full state space are inequivalent, two of their eigenvalues are different. All of these equilibrium points have an unstable direction with eigenvalue $-2\sqrt{3}\Sigma_{23}$ corresponding to the variable $A$. 
These equilibrium points are given in \cite{HHC}. 
\subsubsection{$T(VI_h)$: $D=0$ case ($\Omega>0$, $V<1$)}
\begin{enumerate}
\item{} $\mathcal{C}(h)$: Collins perfect fluid solutions, $h<0$, $2/3<\gamma<\frac{2(1-h)}{1-3h}$ \\
$\Sigma_{12}=\Sigma_{13}=\Sigma_{23}=\lambda=V=0$, $\Sigma_+=-\frac 1{4}(3\gamma-2)$, $\Sigma_-=\frac{\sqrt{-3h}}{4}(3\gamma-2)$, $N^2=\frac 3{16}(3\gamma-2)(2-\gamma)$, $A=\sqrt{-3h}N$, $\Omega=\frac 34\left[(2-\gamma)+h(3\gamma-2)\right]$. 

This equilibrium point is in $B(VI_h)$. 

Eigenvalues: 
\beq &&\lambda_1+\lambda_2=\lambda_3+\lambda_4=-\frac 32(2-\gamma), \quad \lambda_1\lambda_2>0,~\lambda_3\lambda_4>0,\nonumber \\
&& \lambda_5=-\frac{(2-\gamma)}{\gamma}, \quad \lambda_{6,7}=\frac 1{2\gamma}[(5\gamma-6)\pm\sqrt{-h}(3\gamma-2)] \nonumber 
\eeq

\item{} $\mathcal{R}^+(h)$:$\left[\frac{2(3-\sqrt{-h})}{5-3\sqrt{-h}}<\gamma< \frac 32, ~-\frac 19\leq h\leq 0\right]$ and 
$\left[\frac{2(3-\sqrt{-h})}{5-3\sqrt{-h}}<\gamma< \frac{2(1+3\sqrt{-h})}{1+5\sqrt{-h}}, ~ -\frac 14< h< -\frac 19\right]$, $k=\sqrt{-h}$: \\ 
$\Sigma_{13}=\Sigma_{23}=\lambda=v_3=0$, $\Sigma_+=\frac{-b-\sqrt{\Delta}}{2a}$, where 
\begin{multline}\nonumber
\qquad \Delta  = 576\,k^{2}\,\gamma ^{2}\,(\gamma  - 1)\,(1 + 3\,k)^{2}
( - 36\,k^{2} - 24\,k - 4 + 120\,k^{2}\,\gamma  + 160\,k\,\gamma
 + 40\,\gamma  - 133\,k^{2}\,\gamma ^{2} \\
\mbox{} - 182\,k\,\gamma ^{2} - 37\,\gamma ^{2} + 49\,k^{2}\,
\gamma ^{3} + 54\,k\,\gamma ^{3} + 9\,\gamma ^{3})(6 - 5\,\gamma
 + 3\,k\,\gamma  - 2\,k)^{2} 
\end{multline}
\begin{multline}\nonumber 
 \qquad a= 16\,(35\,\gamma  - 36)\,(\gamma  - 2) + ( - 2880\,\gamma ^{3
} + 12192\,\gamma ^{2} - 15936\,\gamma  + 6528)\,k +  \\
(33792\,\gamma ^{2} + 8064 - 18432\,\gamma ^{3} - 26880\,\gamma
 + 3600\,\gamma ^{4})\,k^{2} -  \\
288\,(\gamma  - 1)\,(15\,\gamma ^{3} - 47\,\gamma ^{2} + 46\,
\gamma  - 12)\,k^{3} + 432\,\gamma \,(3\,\gamma  - 2)\,(\gamma
 - 1)^{2}\,k^{4}
\end{multline}
\begin{multline}\nonumber 
\qquad b= 8\,(\gamma  - 2)\,(35\,\gamma  - 36)\,(3\,\gamma  - 2) + (
 - 6528 + 11976\,\gamma ^{3} + 20256\,\gamma  - 2160\,\gamma ^{4}
 - 23616\,\gamma ^{2})\,k \\
 + (14280\,\gamma ^{3} + 25248\,\gamma  - 8064 - 2544\,\gamma ^{4
} - 28800\,\gamma ^{2})\,k^{2} +  \\
( - 10248\,\gamma ^{3} + 19392\,\gamma ^{2} + 1584\,\gamma ^{4}
 - 14112\,\gamma  + 3456)\,k^{3} + 144\,\gamma ^{2}\,(3\,\gamma
 - 2)\,(\gamma  - 1)\,k^{4}.
\end{multline}
$\Sigma_-=\frac{(3\gamma-2)+4\Sigma_++k(1-2\Sigma_+)(5\gamma-6)}{2\sqrt{3}\gamma k}$,  $\Sigma_{12}^2=\frac{(1-2\Sigma_+)(C_0+C_1\Sigma_+)}{6k\gamma^2[(3\gamma-2)k-(5\gamma-6)]}$ where 
\beq
C_0&=&(2-\gamma)(5\gamma-4)(3\gamma-2)+2(-19\gamma^3+60\gamma^2-64\gamma+24)k-\gamma(3\gamma-2)(5\gamma-6)k^2\nonumber \\
C_1&=&4 (2-\gamma)(5\gamma-4)+8(5\gamma^3-24\gamma^2+32\gamma-12)k-4\gamma(3\gamma-2)(2\gamma-3)k^2\nonumber
\eeq
$N^2=\frac{\left[(3\gamma-2)+4\Sigma_+\right]\left[4(5\gamma-6)(3k(\gamma-1)-1)\Sigma_+-(3\gamma-2)(5\gamma-6)+3k(7\gamma-6)(2-\gamma)\right]}{12k^2\gamma^2[(3\gamma-2)k-(5\gamma-6)]}$, $A=\sqrt{3}kN$. The
expressions for $\Omega$ and $V^2$ are given in the appendix, eqs.(\ref{OmegaC4}) and (\ref{VsqC4}).

This equilibrium point lies in $T^+_2(VI_h)$.

\paragraph{Eigenvalues:}
\[ \re(\lambda_{1,2,3,4})<0, \quad \lambda_{5}+\lambda_6=-2(1+\Sigma_+),\quad  \lambda_5\lambda_6>0, \quad\lambda_7=\frac 12(5\gamma-6)(3-2Av_1).\]
This equilibrium point is stable in $N^+(VI_h)$, but always has one unstable eigenvalue in $T(VI_h)$. 

\item{} $\mathcal{R}^-(h)$: $\left[\frac{2(3+\sqrt{-h})}{5+3\sqrt{-h}}<\gamma<\frac 32, ~-\frac 19\leq h\leq 0\right]$ and 
$\left[1<\gamma<\frac{2(3+\sqrt{-h})}{5+3\sqrt{-h}},~ -1< h< -\frac 19\right]$, $k=-\sqrt{-h}$: \\
This point is given via the expressions for $\mathcal{R}^+(h)$ above with $k<0$ but $\Sigma_+=\frac{-b+\sqrt{\Delta}}{2a}$ and using the symmetries $\phi_2$, $\phi_1$ and $\phi_4$ (thus interchanging $\Sigma_{12}$ and $\Sigma_{13}$, and $v_2$ and $v_3$ so that $\Sigma_{12}=v_2=0$). 

This equilibrium point lies in $T^-_2(VI_h)$. 

\paragraph{Eigenvalues:}
\beq &&\re(\lambda_{1,2,3})<0, \quad \re(\lambda_4)=\begin{cases} <0, & -1/9<h\leq 0 \\ >0, & -1<h<-1/9, \end{cases} \nonumber \\ 
&& \lambda_{5}+\lambda_6=-2(1+\Sigma_+),\quad \lambda_5\lambda_6>0, \quad\lambda_7=\frac 12(5\gamma-6)(3-2Av_1).\nonumber \eeq
This equilibrium point is stable in $N^-(VI_h)$ for $-1/9<h\leq 0$, is stable in $T(VI_h)$ for $\frac{2(3+\sqrt{-h})}{5+3\sqrt{-h}}<\gamma<\frac 65, ~-\frac 19\leq h\leq 0$ and has one unstable eigenvalue otherwise. 
The eigenvalue $\lambda_4$ corresponds to a direction in $T^-(VI_h)$, while $\lambda_7$ corresponds to the quantity $D$, defined earlier. 

\end{enumerate}

\subsubsection{$T(VI_h)$: Vacuum case ($\Omega=0$)} \label{pp}
All of these equilibrium points are plane wave solutions and exist for $0<\gamma<2$, $h<0$. Moreover, they all have $D=0$, and 
\[ \Omega=\Sigma_{12}=\Sigma_{13}=\Sigma_{23}=0, ~\Sigma_-=N=\sqrt{-\Sigma_+(1+\Sigma_+)}, ~A=(1+\Sigma_+),~ -1<\Sigma_+<0,~ |\lambda|<1.\] 
The group parameter is given by $3h\Sigma_+(1-\lambda^2)=(1+\Sigma_+)$. It is also advantageous to introduce $r\equiv\sqrt{1-\lambda^2}$ and the parameter $K^2=-1/h$ ($K$ can have either sign). This implies that we can write 
\[ \Sigma_+=-\frac{K^2}{K^2+3r^2}, \quad 0<r\leq 1. \]  
We will also define $\rho$ by 
\[ \rho=v_2^2+v_3^2.\]

For all equilibrium points with $\Omega=0$, three of the eigenvalues are always 
\[ \lambda_1=0,\quad \lambda_{2,3}=-2[(1+\Sigma_+)\pm 2i\sqrt{3}\lambda N].\]

The equilibrium points are then determined by the tilt velocities:
\begin{enumerate}
\item{} $\mathcal{L}(h)$: $v_1=v_2=v_3=0$. These represent 'non-tilted' plane waves and lie in $B(VI_h)$.

Eigenvalues: 
\beq
&&\lambda_4=-\frac{3}{K^2+3r^2}\left[\gamma(K^2+3r^2)-2(K^2+r^2)\right], \quad \lambda_5=-\frac{3}{K^2+3r^2}\left[2(K^2+2r^2)-\gamma(K^2+3r^2)\right], \nonumber \\
&& \lambda_{6,7}=-\frac{3}{K^2+3r^2}\left[(K^2+4r^2\pm |K|r^2)-\gamma(K^2+3r^2)\right] \nonumber
\eeq
\item{} $\widetilde{\mathcal{L}}(h)$: $v_1=\frac{\gamma(K^2+3r^2)-2(K^2+2r^2)}{2r^2(\gamma-1)}$, $v_2=v_3=0$, $\frac{2(K^2+3r^2)}{K^2+5r^2}<\gamma<2$. These represent 'intermediately tilted' plane waves and lie in $T_1(VI_h)$. 

Eigenvalues: 
\beq
&&\lambda_4=-\frac{3(K^2+r^2)(2-\gamma)}{(K^2+3r^2)(\gamma-1)}, \nonumber \\ && \lambda_5=-\frac{3(K^2+r^2)[(K^2+5r^2)\gamma-2(K^2+3r^2)][(K^2+3r^2)\gamma-2(K^2+2r^2)]}{(K^2+3r^2)(\gamma-1)[(K^2+3r^2)^2\gamma-2(K^4+4K^2r^2+5r^4)]}\nonumber \\
&&\lambda_{6,7}=-\frac{3(K^2+r^2)[3\gamma-4\pm|K|(2-\gamma)]}{2(K^2+3r^2)(\gamma-1)}\nonumber
\eeq

\item{} $\widetilde{\mathcal{L}}_{\pm}(h)$: $v_1=\pm 1$, $v_2=v_3=0$. These represent 'extremely tilted' plane waves and lie in $T_1(VI_h)$. 

Eigenvalues: 
\beq \widetilde{\mathcal{L}}_{+}(h): && \lambda_4=0,\quad \lambda_5=2\lambda_6=2\lambda_7=\frac{6(K^2+r^2)}{(K^2+3r^2)}\nonumber \\
\widetilde{\mathcal{L}}_{-}(h): && \lambda_4=-\frac{12r^2}{K^2+3r^2}, \quad \lambda_5=-\frac{6[(K^2+5r^2)\gamma-2(K^2+3r^2)]}{(K^2+3r^2)(2-\gamma)}, \nonumber \\
&& \lambda_{6,7}=-\frac{3}{K^2+3r^2}(r^2\pm 2|K|r^2-K^2)\nonumber.
\eeq
\item{}\label{defF} $\widetilde{\mathcal{F}}^+(h)$: Here $K>0$ and
\beq v_1&=&-\frac{\gamma(K^2+3r^2)-(K^2+Kr^2+4r^2)}{r^2(3-2\gamma+K)},\quad v_2^2-v_3^2=\mathrm{sign}(K) \rho r \nonumber \\  \rho&=&\frac{(K^2+r^2)\left[2(2+K)-(3+K)\gamma\right]\left[\gamma(K^2+3r^2)-(K^2+Kr^2+4r^2)\right]}{r^4(1+K)(3-2\gamma+K)^2} \nonumber
\eeq
where 
\beq
\frac{2(K+2)}{K+3}\leq \gamma\leq \frac{K^2+(4+K)r^2}{K^2+3r^2}, & \text{for} & 0<K<r^2, \nonumber \\
\frac{K^2+(4+K)r^2}{K^2+3r^2}\leq \gamma \leq \frac{2(K+2)}{K+3} , & \text{for} & r^2\leq K<r(r+\sqrt{r^2+1}), \nonumber \\
\frac{K^2+(4+K)r^2}{K^2+3r^2}\leq \gamma \leq \frac{K^2+3r^2}{K^2+(2-K)r^2} , & \text{for} & r(r+\sqrt{r^2+1})\leq K, \nonumber
\eeq
These represent 'intermediately tilted' plane waves and lie in $N^+(VI_h)$ (for $\lambda=0$ they lie in $T_2^+(VI_h)$).  

Eigenvalues: 
\beq &&\lambda_4=-\frac{3(K^2+r^2)[(5+K)\gamma-2(3+K)]}{(K^2+3r^2)(3-2\gamma+K)}, \quad \lambda_5=2\sqrt{3}(1-v_1)\mathrm{sign}(K)Nr, \nonumber \\
&& \lambda_6+\lambda_7=-\frac{3(K^2+r^2)F(K,r,\gamma)}{(3-2\gamma+K)(K^2+3r^2)G(K,r,\gamma)}, \nonumber \\
&& \lambda_6\lambda_7=\frac{18(K^2+r^2)^2\left[2(2+K)-(3+K)\gamma\right]}{(K^2+3r^2)^2(3-2\gamma+K)G(K,r,\gamma)}\nonumber \\
&& \phantom{\lambda_6\lambda_7=}\quad \times\left[\gamma(K^2+3r^2)-(K^2+Kr^2+4r^2)\right]\left[(K^2+3r^2)-\gamma(K^2+(2-K)r^2)\right], \nonumber 
\eeq
where we have defined 
\beq
 && G(K,r,\gamma)= ( - K^{4} + 3r^{4} - 4K^{2}r^{2} + 2K^{2}r^{4} + 8r^{
4}K)\nonumber \\ 
&& \phantom{G(K,r,\gamma)=}+ (7K^{2}r^{2} - 9r^{4}K - K^{2}r^{4} + 2K^{4}
 - K^{3}r^{2} - 2r^{4})\gamma +  K( - K + r^{2})(K^{2} + 3r^{2})\gamma^{2} \nonumber \\
&& F(K,r,\gamma)= 2(K + 3)( - K^{4} - 5K^{2}r^{2} + 2K^{2}r^{4} + 6r
^{4}K - 2r^{4}) \nonumber \\ 
&& \phantom{G(K,r,\gamma)=}+ (16K^{3}r^{2} - 65r^{4}K + 17K^{4} + 78K^{2}r^{2}
 - 38K^{2}r^{4} + 5K^{5} + 41r^{4} - 4r^{4}K^{3} - 2
K^{4}r^{2})\gamma \nonumber \\
&& \phantom{G(K,r,\gamma)=} + ( - 40r^{4} - 67K^{2}r^{2} - 16K^{4} + r^{4}K^{3} +
51r^{4}K + 3K^{4}r^{2} + 20K^{2}r^{4} - 4K^{5} - 4 K^{3}r^{2})\gamma^{2}\nonumber  \\ 
&& \phantom{G(K,r,\gamma)=}
 - (K^{2} + 3r^{2})(K^{2}r^{2} + 5Kr^{2} - 4r^{2} - 5K^{2} - K^{3})\gamma^{3} \nonumber 
\eeq
\item{} $\widetilde{\mathcal{F}}^-(h)$: These are given by the same expressions as for $\widetilde{\mathcal{F}}^+(h)$ but with $ K<0$. The range of $\gamma$ is as follows: 
\beq\frac{K^2+(4+K)r^2}{K^2+3r^2}\leq \gamma \leq \frac{K^2+3r^2}{K^2+(2-K)r^2} , & \text{for} & K\leq r(r-\sqrt{r^2+1}), \nonumber \\
\frac{K^2+(4+K)r^2}{K^2+3r^2}\leq \gamma \leq \frac{2(K+2)}{K+3} , & \text{for} & r(r-\sqrt{r^2+1})\leq K<0, \nonumber 
\eeq
These represent 'intermediately tilted' plane waves and lie in $N^-(VI_h)$ (for $\lambda=0$ they lie in $T_2^-(VI_h)$). 

Eigenvalues: As for $\widetilde{\mathcal{F}}^+(h)$ but with $K<0$.

\item{} $\widetilde{\mathcal{E}}_p^+(h)$: Here $K>r(r+\sqrt{r^2+1})$ and
\beq v_1&=&-\frac{r^2(1+K)}{K(K-r^2)},\quad v_2^2-v_3^2=\mathrm{sign}(K) \rho r \nonumber \\  
\rho&=& 1-v_1^2=\frac{(K^2+r^2)(K^2-2r^2K-r^2)}{K^2(K-r^2)^2}
\nonumber \eeq
These represent 'extremely tilted' plane waves and lie in $N^+(VI_h)$ (for $\lambda=0$ they lie in $T_2^+(VI_h)$).

Eigenvalues: 
\beq
&&\lambda_4=\frac{3(K^2+r^2)[K-r(r+\sqrt{r^2+3})][K-r(r-\sqrt{r^2+3})]}{K(K-r^2)(K^2+3r^2)},\quad \lambda_5=2\sqrt{3}(1-v_1)\mathrm{sign}(K)Nr, \nonumber \\
&&\lambda_6+\lambda_7=-\frac{3(K^2+r^2)[(5r^2+K^2)\gamma-4r^2(2+K)]}{K(K-r^2)(K^2+3r^2)(2-\gamma)}, \nonumber \\
&&\lambda_6\lambda_7=\frac{18(K^2+r^2)^2(K^2-2Kr^2-r^2)[(K^2+2r^2-Kr^2)\gamma-(K^2+3r^2)]}{K^2(K-r^2)^2(K^2+3r^2)^2(2-\gamma)}\nonumber 
\eeq

\item{} $\widetilde{\mathcal{E}}_p^-(h)$: These are given by the same expressions as for $\widetilde{\mathcal{E}}_p^+(h)$ but with $ K<r(r-\sqrt{r^2+1})$.
 
 These represent 'extremely tilted' plane waves and lie in $N^-(VI_h)$ (for $\lambda=0$ they lie in $T_2^-(VI_h)$).
 
 Eigenvalues: As for $\widetilde{\mathcal{E}}_p^+(h)$ but with $ K<r(r-\sqrt{r^2+1})$.
\end{enumerate}

\subsubsection{$T(VI_h)$: $\gamma=6/5$ case ($\Omega>0$, $V<1$, $D\neq 0$)}
\begin{enumerate}
\item{} $\mathcal{B}(h)$, $-1/9<h\leq 0$, $\gamma=6/5$, $k=\sqrt{-h}$:\\
$\Sigma_{23}=0$, $A=\sqrt{3}krN$, $r\equiv\sqrt{1-\lambda^2}$, $v_1=\frac 23A$, $\Sigma_-=\frac 23NA$, $\Sigma_+=-\frac 23(1-A^2)$, \\
$\Omega=\frac{5+V^2}{15}\left[3(1-A^2)-N^2\right]$, $(\Sigma_{12},\Sigma_{13})=\Sigma_{\perp}(\cos\psi,\sin\psi)$, $(v_2,v_3)=v_{\perp}(\cos\phi,\sin\phi)$, \\
$\Sigma_{\perp}^2=-\frac 15V^2(1-A^2)+\frac 13(2+V^2)N^2-\frac 4{25}(1-A^2)^2-\frac 49N^2A^2$, $v_{\perp}^2=V^2-\frac 49A^2$, and $\psi$ is given by $-\pi/2<\psi<0$ and 
\[ \cos(2\psi)=kr\frac{-12{M}N^2+16{M}N^4k^2r^2-15{M}^2v_{\perp}^2+15\Sigma_{\perp}^2N^2r^2(9k^2-1)}{5\Sigma_{\perp}^2\left[2{M}+N^2r^2(1-9k^2)\right]}, \]
where ${M}=\frac 65(1-A^2)-2N^2$.

The tilt velocity $V$ and the angle $\phi$ are given in terms of $(k,N,\lambda)$ (see Appendix, eqs. (\ref{Vsq}) and (\ref{cos2phiC2})). The variable $\lambda$ is determined implicitly by a sixth-order polynomial equation $P(k,N,\lambda)=0$ in $\lambda^2$ where $N$ and $k$ appear as parameters. This polynomial is also given in the appendix, eq. (\ref{Polynomial}). There is a unique solution in terms of $\lambda$, up to symmetries, as long as the range of the parameters $k$ and $N$ are as follows: 
\[ 0\leq k<\frac 13, \qquad N_{\text{lower}}^2\leq N^2\leq \frac{12}{25-9k^2},\]
where 
\[ N_{\text{lower}}^2=\frac{65-60k+27k^2-5(1-3k)\sqrt{9k^2+6k+49}}{(5+3k)(-18k^3-15k^2-20k+25)}.\]
If and only if the parameters $k$ and $N$ obey the bounds above, there exists an equilibrium point for the system. Hence, for every $-1/9<h\leq 0$, $\gamma=6/5$, there exists a one-parameter family of equilibrium points, parameterised by $N$.  In Fig.\ref{P=0} we have plotted the surface $P(k,N,\lambda)=0$ defining the solutions $\mathcal{B}(h)$.
\begin{figure}
\caption{The equilibrium points $\mathcal{B}(h)$: Here the surface $P(k,N,\lambda)=0$ is plotted in $(k,N,r)$-space (recall that $r=\sqrt{1-\lambda^2}$).}\label{P=0}
\centering
\includegraphics[width=9cm]{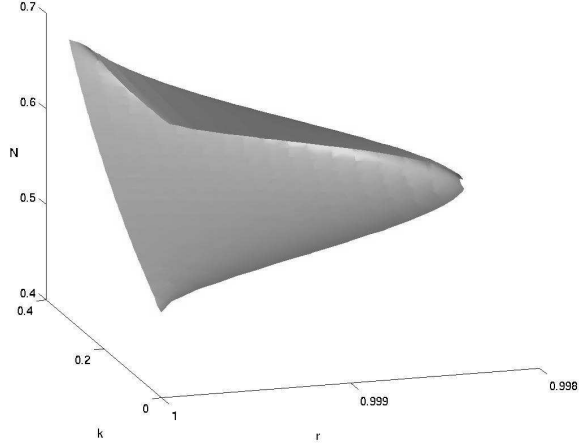} \\
\centerline{\rule[0pt]{9cm}{1pt}}
\end{figure}

This family of non-isolated equilibrium points connects $\mathcal{R}^-(h)$ for $-1/9<h\leq 0$ and $\mathcal{E}_2(h)$ at $\gamma=6/5$. 

\paragraph{Eigenvalues:} Numerical calculation of the eigenvalues shows:
\[ \lambda_1=0, \quad \re(\lambda_{2,3,4,5,6,7})<0.\] 
Hence, this line of equilibria is stable in $T(VI_h)$ whenever it exists.  
\end{enumerate}

\subsubsection{$T(VI_h)$: Extremely tilted case ($\Omega>0$)}
\begin{enumerate} 
\item{} $\mathcal{E}_1^+(h)$, $- 1/9<h\leq 0$ and $0<\gamma<2$, $k=\sqrt{-h}$:\\
$\Sigma_{13}=\Sigma_{23}=v_3=0$, $\Sigma_+=-\frac{(15k^2-2k-5)}{4(1+k)(3k-2)}$, $\Sigma_-=\frac{\sqrt{3}(21k^2-10k-3)}{12(1+k)(3k-2)}$, $\Sigma_{12}=\frac{\sqrt{3-7k^2}(1-3k)}{\sqrt{12}(2-3k)(1+k)}$,\\ $N=\sqrt{\frac{1-k}{2(2-3k)(k+1)^2}}$, $A=\sqrt{3}kN$, $v_1=-\sqrt{\frac{(3-5k)^2}{6(2-3k)(1-k)}}$, $v_2=\sqrt{\frac{(3-7k^2)}{6(2-3k)(1-k)}}$ and $\Omega=\frac{(1-9k^2)(1-k)}{2(2-3k)(1+k)^2}$. 

This equilibrium point lies in $T^+_2(VI_h)$.
\paragraph{Eigenvalues:}
\beq &&\quad \lambda_1=-\frac{(3-7k^2)}{2(2-k-3k^2)}, \quad \lambda_{2,3}=\lambda_{4,5}=-\frac{3(1-k)^2}{4(2-3k)(1+k)}\left[1\pm\sqrt{1-\frac{32(3-7k^2)(1-3k)(2-3k)}{9(1-k)^4}}\right],  \nonumber\\  
&& \quad \lambda_6=-\frac{(3-7k^2)(2\gamma-3)}{(2-3k)(1+k)(2-\gamma)},\quad \lambda_7=\frac{3(3-7k^2)}{2(2-k-3k^2)} \nonumber\eeq
This equilibrium point is stable in $N^+(VI_h)$ for $\gamma>3/2$, and always unstable in $T(VI_h)$. 

\item{} $\mathcal{E}_1^-(h)$, $- 1/9<h\leq 0$ and $0<\gamma<2$, $k=-\sqrt{-h}$:\\
This point is given via the expressions for $\mathcal{E}_1^+(h)$ above with $k<0$ and using the symmetries $\phi_2$, $\phi_1$ and $\phi_4$ (thus interchanging $\Sigma_{12}$ and $\Sigma_{13}$, and $v_2$ and $v_3$ so that $\Sigma_{12}=v_2=0$). 

This  equilibrium point lies in $T^-_2(VI_h)$.

\paragraph{Eigenvalues:} Expressions same as for $\mathcal{E}_1^+(h)$ but with $k<0$.

Stable in $N^-(VI_h)$ for $\gamma>3/2$, always unstable in $T(VI_h)$.

\item{} $\mathcal{E}_2(h)$, $- 1/9<h\leq 0$ and $0<\gamma<2$:\\
$\Sigma_{23}=\lambda=0$, 
$\Sigma_+=-\frac{2(5+9h)}{25+9h}$, $N=\sqrt{\frac{12}{25+9h}}$, $\Sigma_-=\frac 23\sqrt{-3h}N^2$, $A=\sqrt{-3h}N$, $v_1=\frac 23\sqrt{-3h}N$, $\Omega=\frac{6(1+9h)}{25+9h}$, $(\Sigma_{12},\Sigma_{13})=\Sigma_{\perp}(\cos\psi,\sin\psi)$, $(v_2,v_3)=v_{\perp}(\cos\phi,\sin\phi)$  where $\Sigma_{\perp}^2=\frac{15(1+h)(5-27h)}{(25+9h)^2}$, $v_{\perp}^2=\frac{25(1+h)}{25+9h}$, $\cos 2\psi=\frac{27\sqrt{-h}(1+h)}{5-27h}$, $\pi/2<\psi<\pi$, and $\cos 2\phi=\frac{3\sqrt{-h}}{5}$, $0<\phi<\pi/2$.

\paragraph{Eigenvalues:} 
\beq 
&&\lambda_{1,2}=\lambda_{3,4}=-\frac{3(5-3h)}{(25+9h)}\left[1\pm\sqrt{1-\frac{20(1+h)(25+9h)}{(5-3h)^2}}\right]\nonumber \\
&& \lambda_{5,6}=-\frac{15(1+h)}{(25+9h)}\left(1\pm i\sqrt{11}\right), \quad \lambda_7=-\frac{15(1+h)}{25+9h}(5\gamma-6)\nonumber
\eeq
This equilibrium point is stable for $\gamma>6/5$.
\item{} $\mathcal{E}_3(h)$, $- 1<h<-1/9$ and $0<\gamma<2$, $k=-\sqrt{-h}$:\\
$\Sigma_{12}=\Sigma_{23}=\lambda=0$, $\Sigma_+=-\frac{(3k^2+6k-1)}{4(3k-1)}$, $\Sigma_-=-\frac{\sqrt{3}}{4}\frac{(1+6k+k^2)}{1-3k}$, $N=\sqrt{\frac{3(1-k)}{2(1-3k)^2}}$, $A=\sqrt{-3h}N$, $\Sigma_{13}=\sqrt{\frac{3(1-4k-k^2)(1+k)^2}{4(1-3k)^2}}$, $\Omega=-\frac{3(1-k^2)(1+3k)}{2(1-3k)^2}$, $v_1=-\frac{(1+k)}{\sqrt{2(1-k)}}$, $v_2=\sqrt{\frac{1-4k-k^2}{2(1-k)}}$.

This  equilibrium point lies in $T^-_2(VI_h)$.
\paragraph{Eigenvalues:} 
\beq 
&&\lambda_1=-\frac{3(1-4k-k^2)}{2(1-3k)}, \quad\lambda_{2,3}=-\frac{3(1-k)^2}{4(1-3k)}\left[1\pm\sqrt{1-\frac{32(1-4k-k^2)(1+k)}{(1-k)^4}}\right],
\nonumber \\ && \lambda_4=-\frac{3(1-4k-k^2)(\gamma-1)}{(1-3k)(2-\gamma)}, \quad 
\lambda_{5,6}=\lambda_{2,3}, \quad \lambda_7=\lambda_1\nonumber
\eeq
This point is stable for $\gamma>1$. 

\item{} $\mathcal{R}\mathcal{L}(h)$, $0<\gamma<2$, $h<0$:\\
$\Sigma_{12}=\Sigma_{13}=\Sigma_{23}=v_2=v_3=0$, $\Sigma_-=N=(1-\ell)\sqrt{-\Sigma_+(1+\Sigma_+)}$, $A=(1+\Sigma_+)$, $v_1=1$, $-1<\Sigma_+<0$, $0<\ell<1$, $|\lambda|<1$. The group parameter is given by $3h\Sigma_+(1-\lambda^2)(1-\ell)=(1+\Sigma_+)$. 
These equilibrium points correspond to a non-vacuum plane wave with an extremely tilted fluid. 
These equilibrium points lie in $T_1(VI_h)$ and are always unstable. 
\end{enumerate}
\subsection{Special equilibrium points for $h=-1/9$} 
\begin{enumerate}
\item{} $\mathcal{CF}$: The Collinson-French (Robinson-Trautmann) solution is given by: 
\beq
&&\Sigma_+=-\frac 13,\quad \Sigma_-=\frac{1}{3\sqrt{3}}, \quad
\Sigma_{13}=\frac{\sqrt{15}}{9},\quad
N=\frac{1}{\sqrt{2}}, \quad A=\frac{1}{\sqrt{6}},\nonumber \\
&& \Sigma_{12}=\Sigma_{23}=\Omega=\lambda=0. \nonumber
\eeq 
As for the plane waves the
$v_i$-equations decouple and we can again treat these
separately. The special case, $h=-1/9$ leads to the
exact vanishing of one of the constraint equations, and hence the
tilt-velocity can only have 2 independent components (and we set
$v_3=0$).

There are the following equilibrium points associated with the Collinson-French solution:
\begin{enumerate}
\item{}$\mathcal{CF}_0$: $v_1=v_2=0$, $0<\gamma<2$.
\item{}$\widetilde{\mathcal{CF}}_{1\pm}$: $v_1=-\frac{\sqrt{6}(3\gamma-4)}{2(3-\gamma)}$, $v_2=\pm\frac{\sqrt{5(3\gamma-4)(3-2\gamma)}}{\sqrt{2}(3-\gamma)}$, $\frac 43<\gamma<\frac 32$.
\item{} $\widetilde{\mathcal{CF}}_{2}$: $v_1=\frac{\sqrt{6}(9\gamma-14)}{6(\gamma-1)}$, $v_2=0$, $\frac{24-\sqrt{6}}{15}<\gamma<\frac{24+\sqrt{6}}{15}$.
\item{}$\widetilde{\mathcal{ECF}}_{\pm}$:  $v_1=\pm 1$, $v_2=0$, $0<\gamma<2$.
\end{enumerate}
These equilibrium points, and their stability, were studied in \cite{CH2}. 

\item{} $\mathcal{W}$: Wainwright $\gamma=10/9$ solution:
\beq
&&\Sigma_+=-\frac 13,\quad \Sigma_-=\frac{1}{3\sqrt{3}},\quad 0<\Sigma_{13}<\frac{\sqrt{15}}{9},\quad
N=\frac{1}{6}\sqrt{8+54\Sigma_{13}^2}, \quad A=\frac{1}{\sqrt{3}}N,\quad \Omega=\frac 59-3\Sigma_{13}^2,\nonumber \\
&& \Sigma_{12}=\Sigma_{23}=\lambda=V=0.  \nonumber
\eeq 
This line-bifurcation was studied in \cite{BHtilted} and found to be stable in $T(VI_h)$ for $h=-1/9$. 
\end{enumerate}
\subsection{Special equilibrium points for $h=-1$} 
\begin{enumerate}
\item{} $\mathcal{P}(III)$: There is a special line-bifurcation of 'tilted' vacuum solutions for $h=-1$, $\gamma=1$:
\beq
&& \Sigma_+=-\frac 14, \quad \Sigma_-=N=\frac{\sqrt{3}}{4}, \quad A=\frac 34, \nonumber \\
&&\Sigma_{12}=\Sigma_{13}=\Sigma_{23}=\lambda=0, \quad v_1=v_2=0,\quad 0<v_3<1. \nonumber
\eeq
The extreme limit of this equilibrium point, $\lim_{v_3\rightarrow 1}\mathcal{P}(III)=\left.\widetilde{\mathcal{E}}_p^-(h)\right|_{(h,r)=(-1,1)}$. We will call $ \left.\widetilde{\mathcal{E}}_p^-(h)\right|_{(h,r)=(-1,1)}$ for $\mathcal{E}^-_p(III)$ for simplicity.  We will also define  $\mathcal{P}_0(III)\equiv \lim_{v_3\rightarrow 0}\mathcal{P}(III)=\left.{\mathcal{L}}(h)\right|_{(h,r)=(-1,1)}$. These solutions are, in fact, locally rotationally symmetric (LRS). 

Regarding the eigenvalues, there are 3 eigenvalues which are zero and the rest all have a negative real part. In order to resolve the stability properties of these equilibrium points one has to resort to centre manifold theory.
\item{} $\mathcal{P}_0(III)\equiv\left.{\mathcal{L}}(h)\right|_{(h,r)=(-1,1)}$.
\item{} $\mathcal{E}^-_p(III)\equiv \left.\widetilde{\mathcal{E}}_p^-(h)\right|_{(h,r)=(-1,1)}$.
\end{enumerate}

\subsection{New self-similar solutions}
As we have provided a complete catalogue of equilibrium points for the tilted type VI$_h$ model, it is of interest to discuss which of these solutions correspond to new solutions. The physically interesting solutions are the intermediately tilted non-vacuum points which correspond to exact vortic  self-similar solutions of type VI$_h$. Self-similar solutions of type VI$_h$ were studied by Rosquist and Jantzen \cite{rosjan}; however, only some type VI$_0$ solutions were found explicitly. The solutions $\mathcal{R}^-(h)$ appear to have been found by Apostolopoulos in \cite{Apo1}. Moreover, the $h=-1/9$ case of $\mathcal{R}^+(h)$ appears also to be found by Apostolopoulos in \cite{Apo2}. However, the remaining solutions $\mathcal{R}^+(h)$ for $h\neq -1/9$ seem to be new and have not been given previously in the literature. Furthermore, the exact $\gamma=6/5$ solutions corresponding to $\mathcal{B}(h)$ also appear to be new (although the case $h=0$ was given in \cite{ApoVI0,hervik}).

Regarding the extremely tilted non-vacuum equilibrium points, $\mathcal{E}^{\pm}_1(h)$, $\mathcal{E}_2(h)$ and $\mathcal{E}_3(h)$, these also correspond to exact solutions. We see these equilibrium points exist for all $\gamma \in(0,2)$; however for $\gamma\neq 4/3$ the physical interpretation of these is unclear. For $\gamma=4/3$ we can interpret these solutions as Bianchi type VI$_h$ cosmologies containing null radiation.  

The remaining solutions listed are either known or implicitly known \cite{DS1,CH2}.

\section{Late-time behaviour}
\label{sect:late-time}
Let us discuss the different late-time behaviours.
\subsection{The Realm of the plane-waves}
For the type VII$_h$  and the type IV universes, the vacuum plane-waves played a vital role in the future evolution \cite{HHC}. These plane-wave solutions were also shown to possess a wealth of interesting phenomena, like attracting closed orbits and attracting tori. Since the type IV model is the $h\rightarrow -\infty$ limit of type VI$_h$ models, we expect to find that the plane wave solutions play an important role for, at least, some of the type VI$_h$ models. 

In the Bianchi type VI$_h$ case all of the plane-wave equilibrium points reside in the invariant subspaces $N^{\pm}(VI_h)$. It is illustrative to consider the subspaces $T_2^{\pm}(VI_h)\subset N^{\pm}(VI_h)$ (partly because it is easier to interpret; e.g., the figures for $N^{\pm}(VI_h)$ would be 3-dimensional). The equilibrium points in $T^{\pm}_{2}(VI_h)$ are those with $r=1$. 

\begin{figure}
\caption{The regions of stability of the plane-wave solutions in the invariant subspaces $T^{\pm}_2(VI_h)$. Here $K^2=-1/h$ and $K<0$ corresponds to $T^{-}_2(VI_h)$ while $K>0$ corresponds to $T^{+}_2(VI_h)$. The dashed line corresponds to the lower boundary of the region marked $\widetilde{\mathcal{L}}_-(h)$. The loophole is shown for $K>0$ but has been suppressed for $K<0$.}
\label{FigT2}
\centering
\includegraphics[width=7cm]{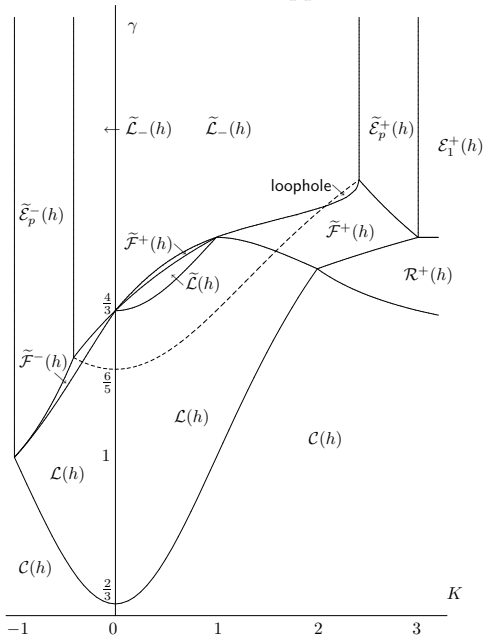}\\
\centerline{\rule[0pt]{9cm}{1pt}}
\end{figure}
In $T^{\pm}_2(VI_h)$ the regions where the various plane-wave equilibrium points are stable are depicted in Fig. \ref{FigT2}. The dashed line indicates the lower bound of stability for $\widetilde{\mathcal{L}}_-(h)$. We can see that there are only stable plane-waves for $h\leq -1$ in $T^-_2(VI_h)$ and $h\leq -1/9$ in $T^+_2(VI_h)$. For the fully tilted models (i.e., in $T(VI_h)$), only the plane-waves in $T^-_2(VI_h)$ remain stable (i.e., the ones with $-1\leq K\leq 0$ in Fig. \ref{FigT2}); all plane waves in $T^+_2(VI_h)$ are unstable in $T(VI_h)$. 
\begin{figure}
\caption{Magnified region of Fig.\ref{FigT2} showing the loophole in $T_2^-(VI_h)$. The curve marked $F=0$ is defined by $F(K,r,\gamma)=0$, where the function $F$ is defined in section \ref{pp}, item (\ref{defF}). Part of this curve marks the threshold value of stability for the equilibrium point $\widetilde{\mathcal{F}}^-(h)$.}
\label{FigT2-}
\centering
\includegraphics[width=6cm]{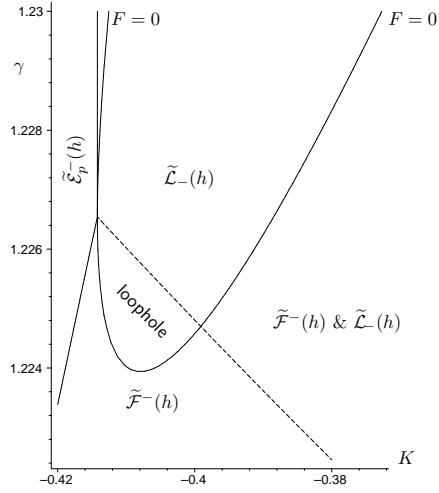}\\
\centerline{\rule[0pt]{9cm}{1pt}}
\end{figure}
\subsubsection{Hopf-bifurcation and the loophole}
The loophole is defined as the set of plane-waves where the variable $\gamma$ satisfies the following: given $K$ and $r=\sqrt{1-\lambda^2}$ where $r(r-\sqrt{r^2+1})<K<r(r+\sqrt{r^2+1})$, then $\gamma_0<\gamma<2(K^2+3r^2)/(K^2+5r^2)$ where $\gamma_0$ is implicitly defined by $F(K,r,\gamma_0)=0$, where $F$ is defined in section \ref{pp},  item (\ref{defF}). 
In the loophole, there are no stable equilibrium points,  so we need to look for other potential candidates for late-time attractors. 

In \cite{CH2,HHC} we noted that the corresponding loopholes in type IV and type VII$_h$ models had closed curves and tori as attractors; hence, we need to look for closed curves in the type VI$_h$ loophole as well. The important observation regarding the nature of the attractor can be seen from the eigenvalues $\lambda_6$ and $\lambda_7$ for the equilibirum point $\widetilde{\mathcal{F}}^{\pm}(h)$. As we vary $\gamma$, we note the following (sufficiently close to $\gamma_0$): 
\begin{itemize}
\item $\gamma<\gamma_0$: $\lambda_6+\lambda_7<0$, $\lambda_6\lambda_7>0$.
\item $\gamma=\gamma_0$: $\lambda_6+\lambda_7=0$, $\lambda_6\lambda_7>0$.
\item $\gamma>\gamma_0$: $\lambda_6+\lambda_7>0$, $\lambda_6\lambda_7>0$.
\end{itemize}
This implies that the real values of $\lambda_{6,7}$ change sign while the imaginary values remain non-zero; this is an indication of a possible \emph{Hopf-bifurcation}. 

Consider a 2-dimensional dynamical system given by the complex variable $Z$. The normal form of a Hopf-bifurcation can be written 
\beq
Z'=(\lambda+b|Z|^2)Z,
\eeq 
where $b$ is some complex number and $\lambda$ is a parameter. If $\re(b)$ is negative then there are stable closed orbits for $\lambda>0$. In order to determine whether our system experiences a Hopf-bifurcation as we vary $K$, $r$ and $\gamma$ we thus need to expand to 3rd order in the variables. This has proven to be difficult in practice due to the complicated dependence on the parameters $K$, $r$ and $\gamma$; however, there is analytic as well as numerical evidence that the fixed points $\widetilde{\mathcal{F}}^{\pm}(h)$ do indeed experience a Hopf-bifurcation and produce a stable closed orbit as $\gamma$ passes through $\gamma_0$ for the following ranges:
\beq
\widetilde{\mathcal{F}}^{+}(h): && r^2<K<r(r+\sqrt{r^2+1}), \nonumber \\
\widetilde{\mathcal{F}}^{-}(h): && r(r-\sqrt{r^2+1})<K<0. \nonumber 
\eeq
This does indicate that there are attracting closed curves even outside the loophole; however, outside the loophole these attracting curves will co-exist with attracting equilibrium points. Inside the loophole, on the other hand, only the closed curves can be attactors as all the equilibrium points are unstable. The family of such attracting closed orbits, regardless whether they are outside or inside the loophole, will be referred to as \emph{the Mussel attractor}. For numerical plots of the Mussel attractor outside the loophole, see Figs. \ref{T2_fig_2} and \ref{FF6}.

Let us study these closed curves in more detail. As in \cite{HHC}, we consider the tilt-equations in a plane-wave background. We utilize the identity 
\[ (v_2^2+v_3^2)^2=(2v_2v_3)^2+(v_2^2-v_3^2)^2,  \]
and define $(x,\rho,\theta)$ by 
\beq
x=v_1, \quad  \rho\equiv v_2^2+v_3^2, \quad 2v_2v_3=\rho\sin\theta,\quad (v_2^2-v_3^2)= \rho\cos\theta.
\eeq
Then we have for the reduced system
\beq
x'&=& (T+2\Sigma^*_+)x-(A^*+\sqrt{3}N^*\cos\theta)\rho, \nonumber \\
\rho'&=& 2\left(T-\Sigma^*_++A^*x-\alpha\cos\theta\right)\rho, \nonumber \\
\theta'&=& 2\alpha(\lambda^*+\sin\theta),
\label{eq:redtilt}\eeq
where $\alpha=\sqrt{3}(1-x)N^*$ and, by use of the discrete symmetry $\phi_4$, we can assume that $0\leq \theta<2\pi$. Furthermore, these variables are bounded by
\beq
0\leq \rho, \quad V^2= x^2+\rho\leq 1, \quad |\lambda^*|<1.
\eeq
An asterisk has been added to the variables to emphasize that these should be thought of as  the limit values for the full system.

For a periodic orbit, $c(\tau)$, with period $T_n$, we introduce the average of a variable $B$: 
\beq
\av{B}\equiv \frac{1}{T_n}\oint_c Bd\tau.
\eeq
We can also say something about the stability of a closed periodic orbit. For example, consider 
the evolution equation for $\Omega$, which we write as $\Omega'=\lambda_{\Omega}\Omega$. Assume that $c(\tau)$ is 
a closed periodic orbit with period $T_n$. Then $\av{\lambda_{\Omega}}$ indicates the stability of the closed curve with respect to the variable $\Omega$. 

We note that $N^{\pm}(VI_h)$ intersect at $\sin\theta=-\lambda^*$, where
\begin{itemize}
\item $N^+(VI_h)$: $\cos\theta=+\sqrt{1-(\lambda^*)^2}=+r$,
\item $N^-(VI_h)$: $\cos\theta=-\sqrt{1-(\lambda^*)^2}=-r$.
\end{itemize} 
It is in these subspaces that the closed curves must exist. We also note that the variable $\theta$ induces transitions $N^+(VI_h)\rightarrow N^-(VI_h)$ which means that in $T(VI_h)$, $N^+(VI_h)$ is unstable. 

\begin{figure}
\caption{Proof of the existence of the Mussel attractor. }
\label{Fig:ProofMussel}
\centering
\includegraphics[height=7cm]{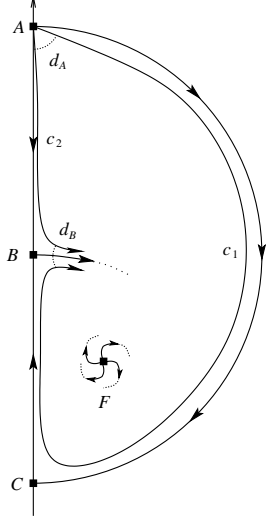}\\
\centerline{\rule[0pt]{9cm}{1pt}}
\end{figure}

\begin{thm}\label{thm:xbar}
Assume that there is a closed properly periodic orbit, $\mathcal{C}_M$ parameterised by $c(\tau)$, for the dynamical system (\ref{eq:redtilt}). Then 
\beq
\av{x}=-\frac{\gamma(K^2+3r^2)-(K^2+Kr^2+4r^2)}{r^2(3-2\gamma+K)}, \quad \av{\lambda_{\Omega}}=-\frac{3(K^2+r^2)[(5+K)\gamma-2(3+K)]}{(K^2+3r^2)(3-2\gamma+K)}.
\eeq
\end{thm} 
\begin{proof}
The proof is analogous to that in the type IV case; see \cite{HHC}.
\end{proof}
Inside the loophole existence can also be proven:
\begin{thm}[Mussel attractor]
For $\lambda^*<1$ and every $K$ and $\gamma$ taking values in the type VI$_h$ loophole  there exists a
closed periodic orbit, $\mathcal{C}_M$,  for the dynamical system (\ref{eq:redtilt}).
\end{thm}
\begin{proof}
The proof is analogous to that in the type IV case \cite{HHC} and goes as follows. Inside the loophole there are no attracting equilibrium points. By restricting to the 2-dimensional set $\cos\theta=\pm r$, we get a picture similar to Fig.\ref{Fig:ProofMussel}. The equilibrium points (labelled $A$, $B$, $C$ and $F$) are all hyperbolic and are sources/saddles as indicated. We note that the boundary consists of heteroclinic orbits (also drawn). We now see that there is an orbit from $B$ which cannot end at an equilibrium point and consequently must approach a closed curve. More explicitly, a future trapping region can be constructed as follows. We can use the shadowing theorem (see Proposition 4.2 in \cite{DS1}) and the fact that the equilibrium points are all hyperbolic to show that there is an orbit, $c_1$, originating from $A$ which can be chosen to come arbitrary close to $C$ and $B$ (also shown in Fig. \ref{Fig:ProofMussel}). Similarly, there is an orbit, $c_2$, originating from $A$ and coming arbitrary close to $B$. We can then cut out neighbourhoods of the points $A$ and $B$ using the dotted curves, $d_A$, $d_B$, so that $c_1$, $c_2$, $d_A$, $d_B$ constitute the boundary of a future trapping region. Since the only equilibrium point inside this trapping region is the source $F$, the Poincar\'e-Bendixson theorem implies the existence of a closed periodic orbit (in particular, there must be an orbit originating from $x=\rho=0$ that approaches a closed orbit).
\end{proof}

There is an important corollary that follows from this: 
\begin{cor}
For $h<-(3-2\sqrt{2})$, there exist closed periodic orbits in $T(VI_h)$.
\end{cor}
The ramifications of this result is that considering the equilibrium points only is not sufficient to determine the asymptotic behaviour for these models. However, apart from the closed periodic orbits described above (the Mussel attractor), we have not found evidence for any other closed period orbits (for $h\neq -1/9$) important for the late-time behaviour.

\subsection{The Realm of non-vacuum vortic Universes} 
\begin{figure}
\caption{The regions of stability of non-vacuum universes in the invariant subspaces $T^{\pm}_2(VI_h)$. Here $k^2=-h$ and $k<0$ corresponds to $T^{-}_2(VI_h)$ while $k>0$ corresponds to $T^{+}_2(VI_h)$. }
\label{FigT2b}
\centering
\includegraphics[width=7cm]{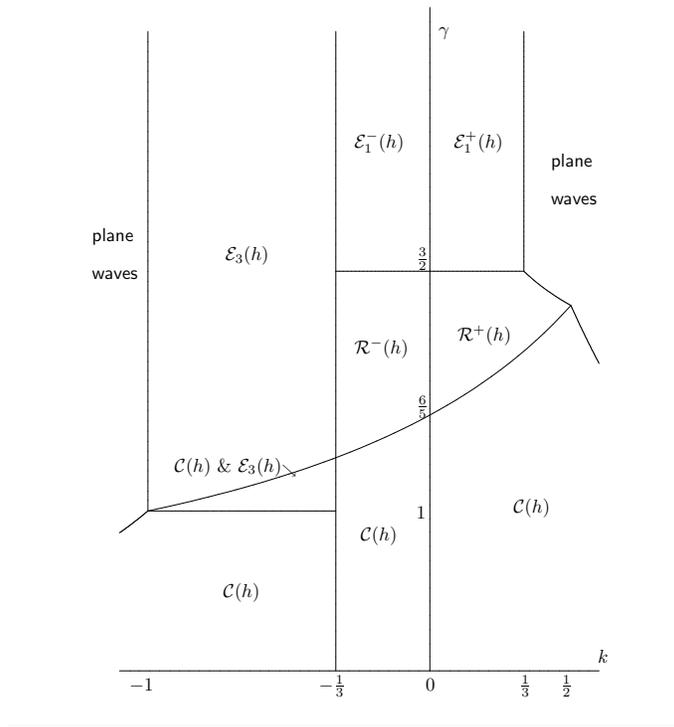}\\
\centerline{\rule[0pt]{9cm}{1pt}}
\end{figure}

We note that all equilibrium points with $\Omega\neq 0$ in $N^{\pm}(VI_h)$ are, in fact, in $T^{\pm}_2(VI_h)$. The regions of stability of the equilibrium points in $T^{\pm}_2(VI_h)$ are depicted in Fig. \ref{FigT2b}. For $N^{\pm}(VI_h)$, the picture is identical. 

\begin{figure}
\caption{The asymptotic behaviour of tilted type VI$_h$ universes. For $h<-1$ we have used the solutions in $T^-_2(VI_h)$ for illustrative purposes only (the figure should be 3-dimensional). Again the loophole has been suppressed.}
\label{FigT2Full}
\centering
\includegraphics[width=7cm]{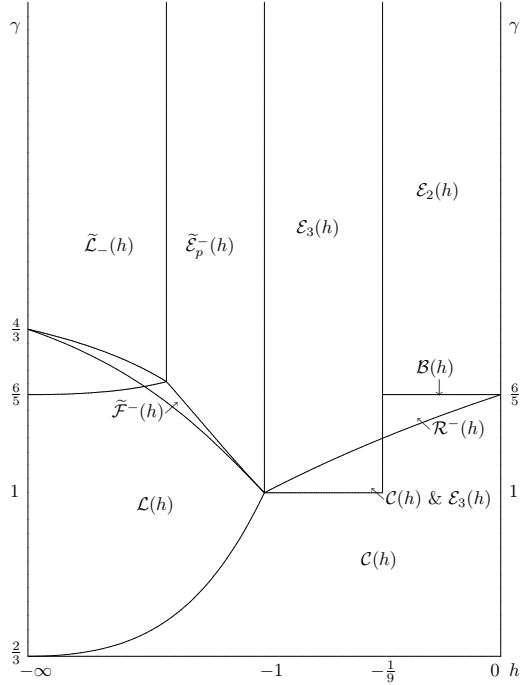}\\
\centerline{\rule[0pt]{9cm}{1pt}}
\end{figure}

In the fully tilted space $T(VI_h)$, we note that from the monotonic function $Z_2$, if $\Omega$ approaches a non-zero value at late times, then $D\rightarrow 0$ and $V\rightarrow 1$ for $\gamma<6/5$ and $\gamma>6/5$, respectively. Our analysis indicates that the future asymptote is non-vacuum for $-1<h<-1/9$ and $-1/9<h\leq 0$, for which we have the following behaviour (see Fig.\ref{FigT2Full}):
\begin{enumerate}
\item{} {{$-1<h<-1/9$:}}
\begin{enumerate}
\item{} $\frac 23<\gamma\leq 1$: $D\rightarrow 0$, and $V\rightarrow 0$. Attractor: $\mathcal{C}(h)$.
\item{} $1<\gamma<\frac{2(3+\sqrt{-h})}{5+3\sqrt{-h}}$: $D\rightarrow 0$, and $V\rightarrow 0$ or $V\rightarrow 1$. Attractor: $\mathcal{C}(h)$ and $\mathcal{E}_3(h)$.
\item{} $\frac{2(3+\sqrt{-h})}{5+3\sqrt{-h}}\leq \gamma <2$: $D\rightarrow 0$, and $V\rightarrow 1$. Attractor $\mathcal{E}_3(h)$.
\end{enumerate}
\item{$-1/9<h\leq 0$:}
\begin{enumerate}
\item{} $\frac 23<\gamma\leq\frac{2(3+\sqrt{-h})}{5+3\sqrt{-h}}$: $D\rightarrow 0$, and $V\rightarrow 0$.  Attractor: $\mathcal{C}(h)$.
\item{} $\frac{2(3+\sqrt{-h})}{5+3\sqrt{-h}}<\gamma<\frac 65$: $D\rightarrow 0$, and $V\rightarrow \mathrm{constant}$.  Attractor: $\mathcal{R}^-(h)$.
\item{} $\gamma =\frac 65$: $D\rightarrow \mathrm{constant}$, and $V\rightarrow \mathrm{constant}$.  Attractor: $\mathcal{B}(h)$.
\item{} $\frac 65<\gamma<2$: $D\rightarrow \mathrm{constant}$, and $V\rightarrow 1$.  Attractor: $\mathcal{E}_2(h)$.
\end{enumerate}\end{enumerate}

\subsection{The case $h=-1$: Bianchi type III}
The case $h=-1$ is a special case which has to be treated separately. For $2/3<\gamma<1$ the Collins solution, $\mathcal{C}(h)$ is the only attractor. All eigenvalues have negative real parts and consequently the stability can be deduced from the linearised analysis. However, for $\gamma=1$ and $\gamma>1$, the equilibrium points $\mathcal{P}(III)$ (including the non-tilted and extreme limits) and $\mathcal{E}^-_p(III)$ have 3 and 2 zero-eigenvalues, respectively. In order to determine the stability for these solutions one must therefore go to higher order. 

Consider first $\gamma=1$. Two of the zero-eigenvalues for $\mathcal{P}(III)$, $0<v_3<1$ correspond to a non-trivial Jordan block: 
\[ J_1=\begin{bmatrix} 0 & 1 \\ 0 & 0 \end{bmatrix}. \] 
This means that the generic solution drifts along the line of equilibria. The solution drifts towards the $v_3=0$ solution, $\mathcal{P}_0(III)$ . It is therefore of interest to study the centre manifold for $\mathcal{P}_0(III)$ . 

\subsubsection{$T_2^-(VI_{-1})$: The centre manifold} 
In our analysis of the centre manifold we restrict ourselves to the invariant subspace $T^-_2(VI_{-1})$. This subspace has only a two-dimensional centre manifold which makes the analysis more tractable and easier for illustrative purposes. It is useful to define the new variables 
\beq
\widehat{\Sigma}_+&=&\frac 12\left(-\Sigma_++\sqrt{3}\Sigma_-\right), \nonumber \\
\widehat{\Sigma}_-&=&\frac 12\left(\sqrt{3}\Sigma_++\Sigma_-\right).
\eeq
The centre manifold can be found as follows. Define $(x,y,z,w)$  by:
\[ (\widehat{\Sigma}_+,N,v_1,v_3)=\left(\frac 12+x, \frac{\sqrt{3}}{4}(1+y),~z,~w\right).\]
The variables $\widehat{\Sigma}_-$, $\Omega$ and $\Sigma_{13}$ can be determined from the contraints. 

By introducing the variables $(X,Y,Z,W)$ as follows: 
\[ (X,Y,Z,W)=\left(x+\frac 32y,x-\frac 32y,-x-\frac 32y+w,-x+\frac 32y+z\right)\]
we can expand the differential equations for $(X,Y,Z,W)$ to second order. The centre manifold can then be seen to be parameterised by $(X,Z)$. Moreover, on the centre manifold $(Y,W)=\mathcal{O}(3)$, where $\mathcal{O}(3)$ means cubic in $X$ and $Z$. The equations for $(X,Z)$ to lowest order are then:
\beq
X' &=& X^2+\mathcal{O}(3), \nonumber \\
Z' &=& XZ+\mathcal{O}(3).
\eeq 
Requiring $\Omega\geq 0$, gives the approximate solutions at late times:
\[ X=-\frac 1{\tau}, \quad Z=\frac{C}{\tau},\]
where $C$ is a constant. 
Substituting these back into the original variables, the decay-rates are found to be:
\beq
\widehat{\Sigma}_+&=&\frac 12\left(1-\frac 1\tau\right), \nonumber \\
\widehat{\Sigma}_-&=&\mathcal{O}\left(\frac{1}{\tau^4}\right), \nonumber \\
\Sigma_{13}&=&\frac{2v_{3,0}}{\sqrt{3}\tau^2}, \nonumber \\
N&=& \frac{\sqrt{3}}4\left(1-\frac 1{3\tau}\right),\nonumber \\
\Omega&=& \frac 1\tau,\nonumber \\
v_1 &=& \mathcal{O}\left(\frac{1}{\tau^3}\right),\nonumber \\
v_3&=& \frac{v_{3,0}}{\tau}.
\eeq
Here, the constant $v_{3,0}$ is related to $C$ by $C=1+v_{3,0}$. 
These decay rates have also been confirmed numerically. We have also simulated the system in the fully tilted space $T(VI_h)$ which does, indeed, confirm that the decay rates have the functional dependence as given above. The numerics also suggest that:
\beq
\lambda=\mathcal{O}\left(\frac{1}{\tau^2}\right), \quad \Sigma_{23}=\mathcal{O}\left(\frac{1}{\tau^3}\right),\quad  v_2=\mathcal{O}\left(\frac{1}{\tau^3}\right), \quad \Sigma_{12}=\mathcal{O}\left(\frac{1}{\tau^4}\right).
\eeq
Hence, it is plausible that: \emph{$\mathcal{P}_0(III)$ is the attractor for the fully tilted dust ($\gamma=1$) type III model.}

For $1<\gamma<2$ the situation is easier. Again the the most promising candidate possesses zero eigenvalues; however, we note that 
\[ \widetilde{\mathcal{E}}_p^-(III)\equiv\lim_{h\rightarrow -1^-}\widetilde{\mathcal{E}}_p^-(h)=\lim_{h\rightarrow -1^+}\mathcal{E}_3(h).\]
Also, for $1<\gamma<2$, $\widetilde{\mathcal{E}}_p^-(h)$,  $\mathcal{E}_3(h)$ are the attractors for $h<-1$ and $h>-1$ (sufficiently close to $-1$), respectively; hence, we would anticipate that  $\widetilde{\mathcal{E}}_p^-(III)$ is the attractor for $h=-1$ and $1<\gamma<2$. Indeed, doing a centre manifold analysis for the subspace $T^-_2(VI_{-1})$ we  get: 
\beq
\widehat{\Sigma}_+=\frac 12 +\mathcal{O}\left(\frac 1{\tau}\right), \quad
N= \frac{\sqrt{3}}4+\mathcal{O}\left(\frac 1{\tau^2}\right),\nonumber \\
v_1=\mathcal{O}\left(\frac 1{\tau}\right), \quad 
v_3=1+\mathcal{O}\left(\frac{1}{\tau^2}\right), \quad 
\Omega=\mathcal{O}\left(\frac 1{\tau}\right).
\eeq
This confirms that the type III models with $1<\gamma<2$ at late times approach the extremely tilted vacuum solution $\widetilde{\mathcal{E}}_p^-(III)$.

Note that we have only presented here a centre manifold analysis for the invariant subspace $T^-(VI_{-1})$. A full analysis of these centre manifolds have been performed showing that these equilibrium points are indeed the late-time attactors; however, due to the lengthy and complicated nature of the centre manifold analysis, this analysis will be presented elsewhere. 

There is an interesting connection  with the type VIII models \cite{HLim} worth pointing out. We note that the type III solutions approach an LRS universe at late times. In fact, type III LRS solutions also admit a type VIII action. Comparing the attractors for type III and type VIII we note there is a striking similarity: In terms of the metric and the matter (in a $C^0$ sense), the type III and type VIII attractors are the same. Explicitly, the curvature $A^2+N^2$ in the type III model takes the role of $-N_1\bar{N}$ in the type VIII model. In fact, by comparing the decay rates we even see that some of the decay rates are identical to lowest order. On the other hand, the two models do approach this attractor differently since the Weyl parameter diverges for type VIII while it is always bounded for type III.   

\subsection{The case $h=-1/9$} 
Based on the eigenvalues of the equilibrium points, the equilibrium points are future attractors in the range specified \cite{CH2}:
\begin{enumerate}
\item{}$\mathcal{CF}_0$: $10/9<\gamma\leq 4/3$.
\item{}$\widetilde{\mathcal{CF}}_{1\pm}$: $\frac 43<\gamma<\frac 56+\frac{\sqrt{721}}{42}$.
\item{} $\widetilde{\mathcal{CF}}_{2}$: Never an attractor.
\item{}$\widetilde{\mathcal{ECF}}_{-}$: $\frac{24-\sqrt{6}}{15}<\gamma<2$.
\end{enumerate}

Including the analysis of the non-tilted equilibrium
points we can thus conclude:
\begin{itemize}
\item{} $2/3<\gamma<10/9$: Collins type VI$_{-1/9}$ perfect fluid solution is stable.
\item{} $\gamma=10/9$: Wainwright's type VI$^*_{-1/9}$, $\gamma=10/9$, solution is stable.
\item{} $10/9<\gamma<2$: The Collinson-French vacuum is stable. Moreover, for $4/3<\gamma$ the asymptotic tilt velocity is non-zero.
\end{itemize}
It should be emphasised that the analysis of this model is not quite complete. Due to the exact vanishing of one of the contraint equations a different approach has to be taken regarding the numerical analysis.  Furthermore, the question whether there are closed periodic orbits acting as attractors has not been addressed in this paper. Preliminary analysis indicates that this model also possesses closed periodic orbits \cite{HHLC19}.

\section{Numerical Analysis}
\label{sect:num}
\subsection{Additional Numerical Support}
In the qualitative analysis of the system of equations governing the
evolution of the tilted Bianchi type VI$_h$ models, it was found that
there were numerous possible future asymptotic states.  In addition,
it was also found that in the invariant set $T_2^+(VI_h)$ there is a
portion of the parameter space,  \emph{the loophole}, in which there is
no equilibrium point acting as a future asymptotic attractor.  An
extensive numerical analysis has been completed that further supports
the qualitative analysis presented in the paper.  In addition to
obtaining numerical evidence for all behaviours for the general type
VI$_h$, and in the special case of the invariant set $T_2^+(VI_h)$,
some very special and even somewhat surprising results have been
found.  Below we include a select group of numerical plots that
illustrate some of the rich behaviour found in the type VI$_h$
models.  More details and numerical plots can be found in the
companion article \cite{companion2007}.

\subsection{The invariant set $T_2^+(VI_h)$.}
 
 Here, we are looking at the dynamics in the four-dimensional
invariant set $T_2^+(VI_h)$. (Recall that
$\Sigma_{13}=\Sigma_{23}=\lambda=v_3=0$, and $A=\sqrt{-3h}N$ in the
$T_2^+(VI_h)$ invariant set.)  A six-dimensional set of differential
equations for the variables
$\{\Sigma_{+},\Sigma_{-},\Sigma_{12},N,v_1,v_2\}$ was integrated
using typical values for the parameters $h$ and $\gamma$ that
illustrate some different asymptotic behaviours.  The constraint
equations were used to check the accuracy of the numerical
integrations.  The built in differential equations solver {\em ode45}
in Matlab R2006a has been used to numerically solve all cases.  The
{\em Relative Tolerance} was set to $10^{-12}$ and the {\em Absolute
Tolerance} was set to $10^{-20}$ or smaller.  The constraint
equations were used to determine the initial values of the tilt
velocities $v_1, v_2$.
 
\begin{figure}[f]
\caption{The invariant set $T_2^+(VI_h)$: $h=-\frac{1}{5}, \gamma=1.59$:  Attractor is the Mussel Attractor.  Initial Conditions were chosen to be $\Sigma_{+}=3/10 - i/10, \Sigma_{-}=-0.3+i/10,\Sigma_{12}=4/10,N=i/20$ where i ranges from 1 to 4.}\label{T2_fig_1}
$$\includegraphics[scale=0.5,angle=90]{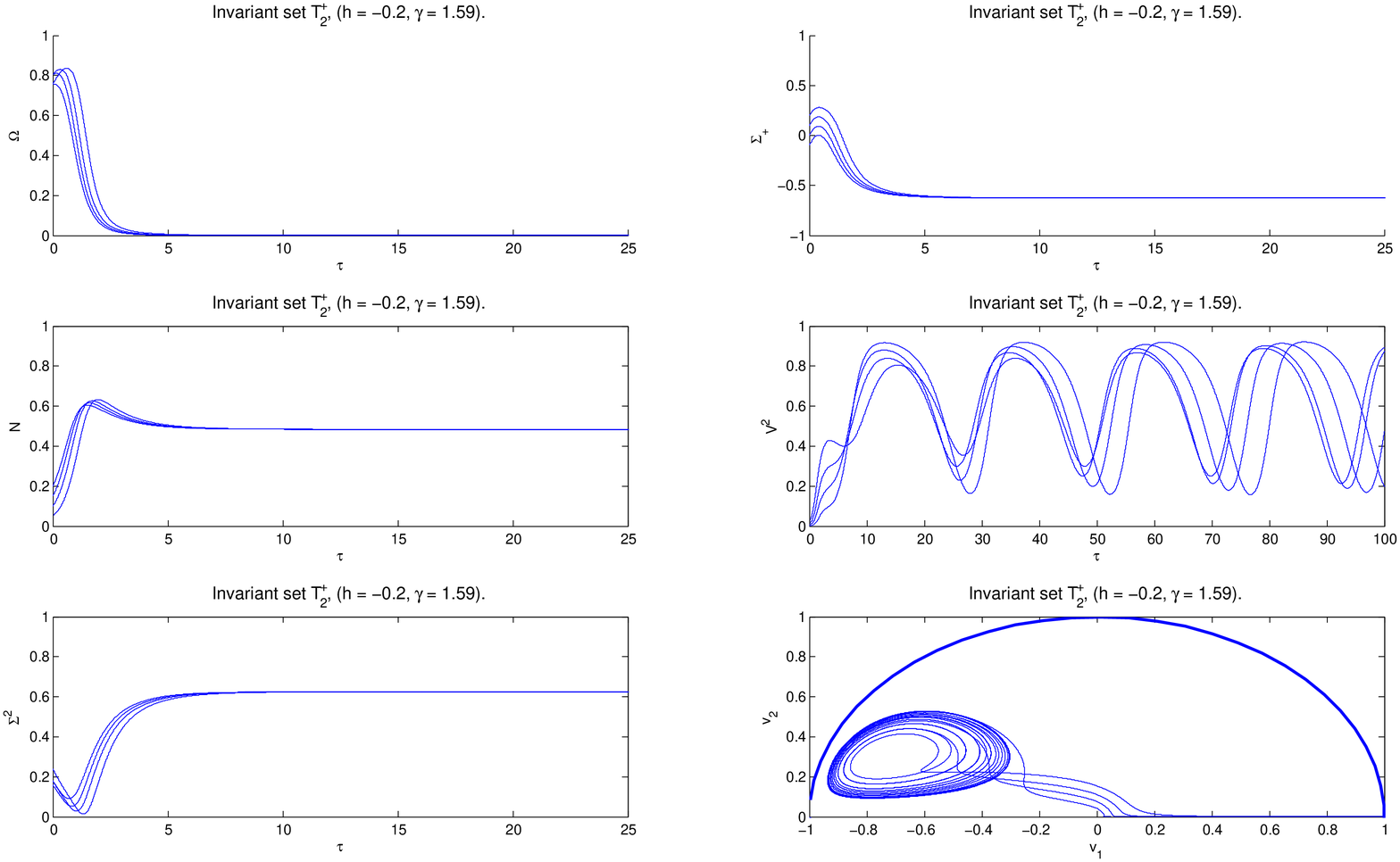}$$
\end{figure}

\begin{figure}[f]
\caption{The invariant set $T_2^+(VI_h)$: $h=-\frac{13}{50}, \gamma=1.565$:  Attractors are $\tilde{\mathcal L}_{-}(h)$ and the Mussel attractor(in red): Note these parameter values are not in the loophole.  Initial Conditions were chosen to be $\Sigma_{+}=3/10 - i/10, \Sigma_{-}=-0.3+i/10,\Sigma_{12}=4/10,N=i/10$}\label{T2_fig_2}
$$\includegraphics[scale=0.5,angle=90]{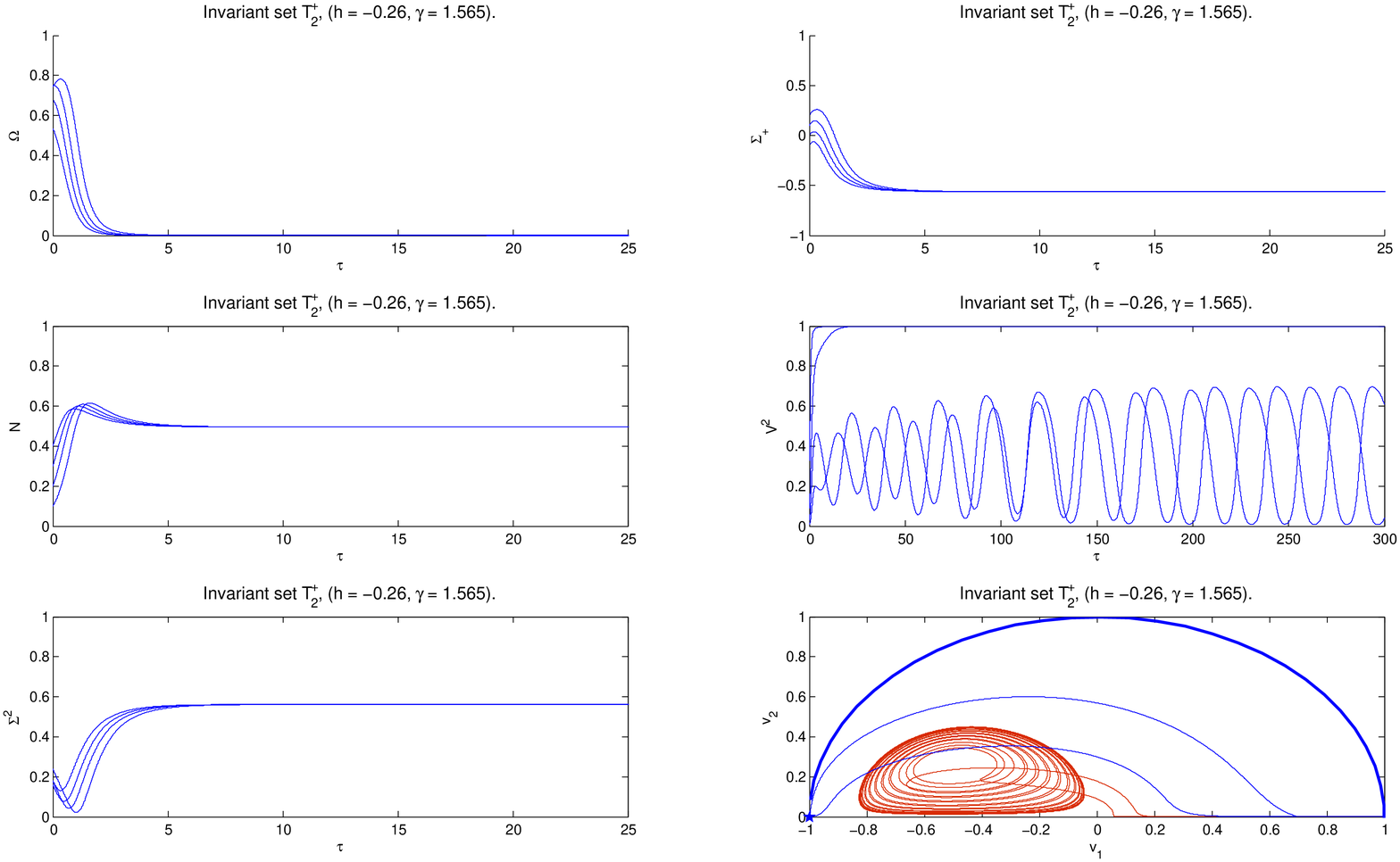}$$
\end{figure}

Two numerical plots, Figures \ref{T2_fig_1} and \ref{T2_fig_2} are
included here that show some of the more surprising (albeit not
totally unexpected) behaviour in the invariant set $T_2^+(VI_h)$.
Note the figures included here do not show typical behaviour for the
entire range of parameter space for $T_2^+(VI_h)$, but only
illustrate what is typical for small neighborhoods around the
selected parameter values.  Indeed other behaviours are possible;
there are additional situations when the future attractor is not
unique, although, no others show evidence of closed orbits.  For
additional illustrations see the companion article \cite{companion2007}.

In Figure \ref{T2_fig_1} we observe four periods of the variable
$V^2$ and the corresponding Mussel attractor in the accompanying
$(v_1,v_2)$ phase portrait.  The existence of this closed orbit was
predicted, since the lack of a stable equilibrium point inside the
loophole and the previous analysis of the Bianchi type VII$_h$ and
IV suggested that it must be true for at least some class of the
type VI$_h$ models.  However, what was not predicted before the
numerical analysis was completed was the existence of a closed orbit
for parameter values outside the loophole.  Figure \ref{T2_fig_2}
shows that it is indeed possible to have a Mussel attractor for
parameter values outside the loophole.  In addition, in this
situation it can be seen that there are two attractors, the Mussel
attractor which shows a periodic nature in the variable $V^2$ and the
equilibrium point $\widetilde{\mathcal L}_{-}(h)$ which is extremely
tilted.

\subsection{Fully tilted Bianchi VI$_{h}$ space.}
 
\begin{figure}[f]
\caption{Fully tilted Bianchi VI$_{h}$: $h=-\frac{1}{16}, \gamma=1.20$:  The initial conditions are chosen to be $\Sigma_{+}=-\Sigma_{-}=0.3-(15+i)/40,N= (15+i)/40,\lambda=0.6- (15+i)/20$ with $i$ ranging from 1 to 5.  Note the non-isolated nature of the line of equilibria ${\mathcal B}(h)$. }\label{FA4}
$$\includegraphics[scale=0.5,angle=90]{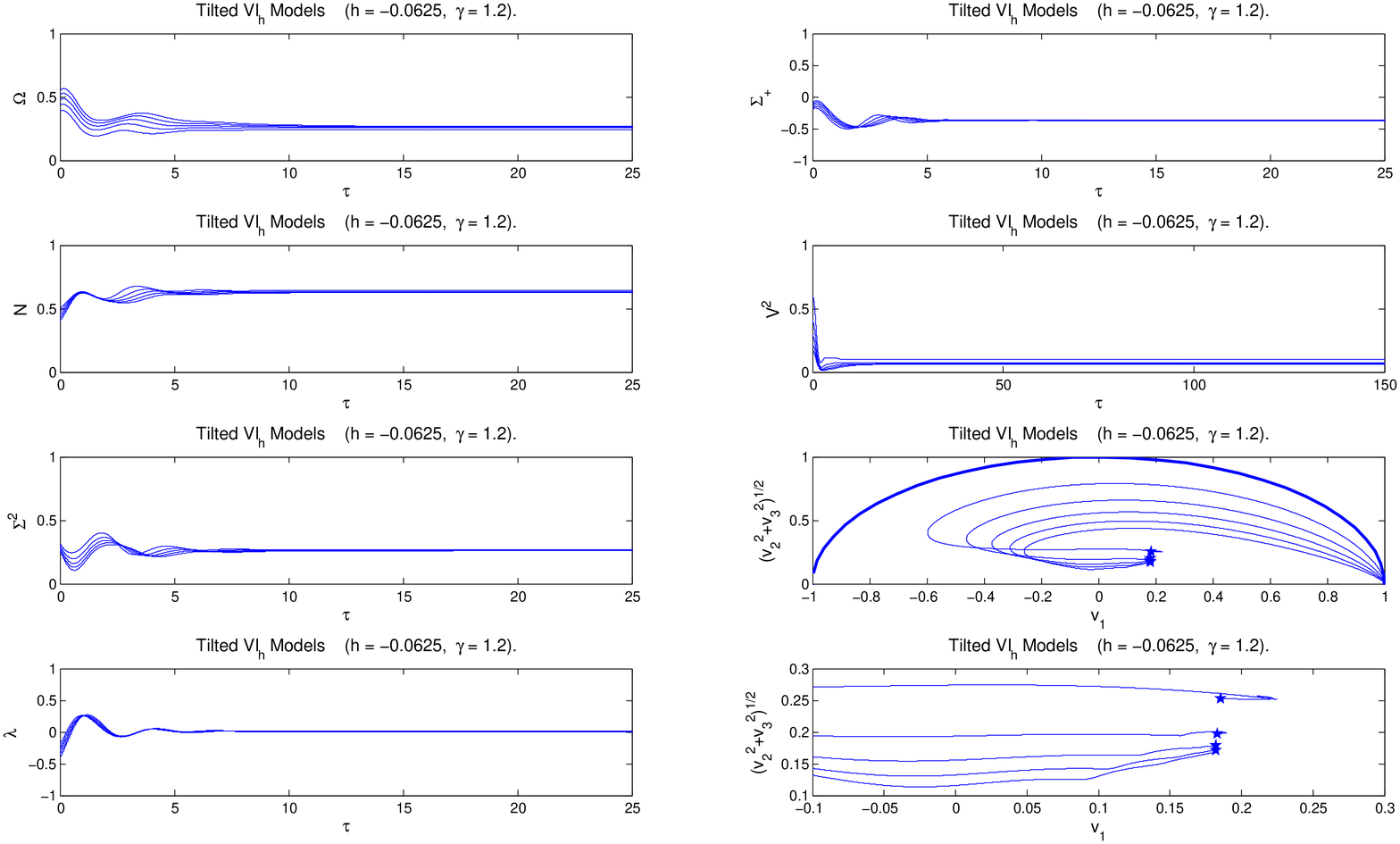}$$
\end{figure}

\begin{figure}[f]
\caption{Fully tilted Bianchi VI$_{h}$: $h=-5/9, \gamma=1.035$:  Attractors are ${\mathcal C}(h)$ and ${\mathcal E}_3(h)$. The initial conditions are chosen to be $\Sigma_{+}= 0.3-\frac{i}{10},\Sigma_{-}= -0.3+\frac{i}{10},\Sigma_{12}=0.4,\Sigma_{13}=0.2,\Sigma_{23}=0.2,N= \frac{i}{20},\lambda=0.6-\frac{i}{5}$ with $i$ ranging from 1 to 6.  Note the slow transition which is due to a relatively small eigenvalue $\lambda\sim 0.02$ of the unstable equilibrium point $\mathcal{R}^-(h)$. }\label{FC2}
$$\includegraphics[scale=0.5,angle=90]{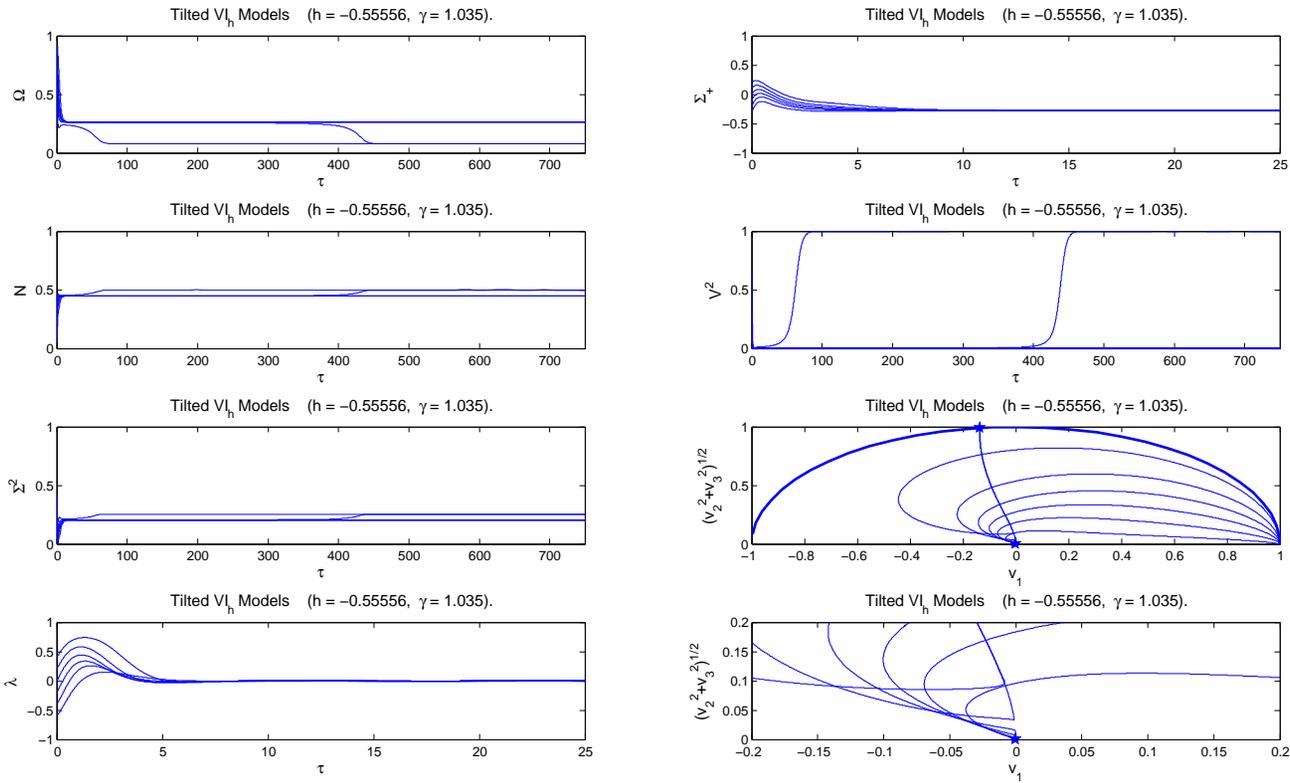}$$
\end{figure}

\begin{figure}[f]
\caption{Fully tilted Bianchi VI$_{h}$: $h=-5/9, \gamma=1.10$:  Attractor is ${\mathcal E}_3(h)$. The initial conditions are chosen to be $\Sigma_{+}= 0.3-\frac{i}{10},\Sigma_{-}= -0.3+\frac{i}{10},\Sigma_{12}=0.4,\Sigma_{13}=0.2,\Sigma_{23}=0.2,N= \frac{i}{20},\lambda=0.6-\frac{i}{5}$ with $i$ ranging from 1 to 5.}\label{FC3}
$$\includegraphics[scale=0.5,angle=90]{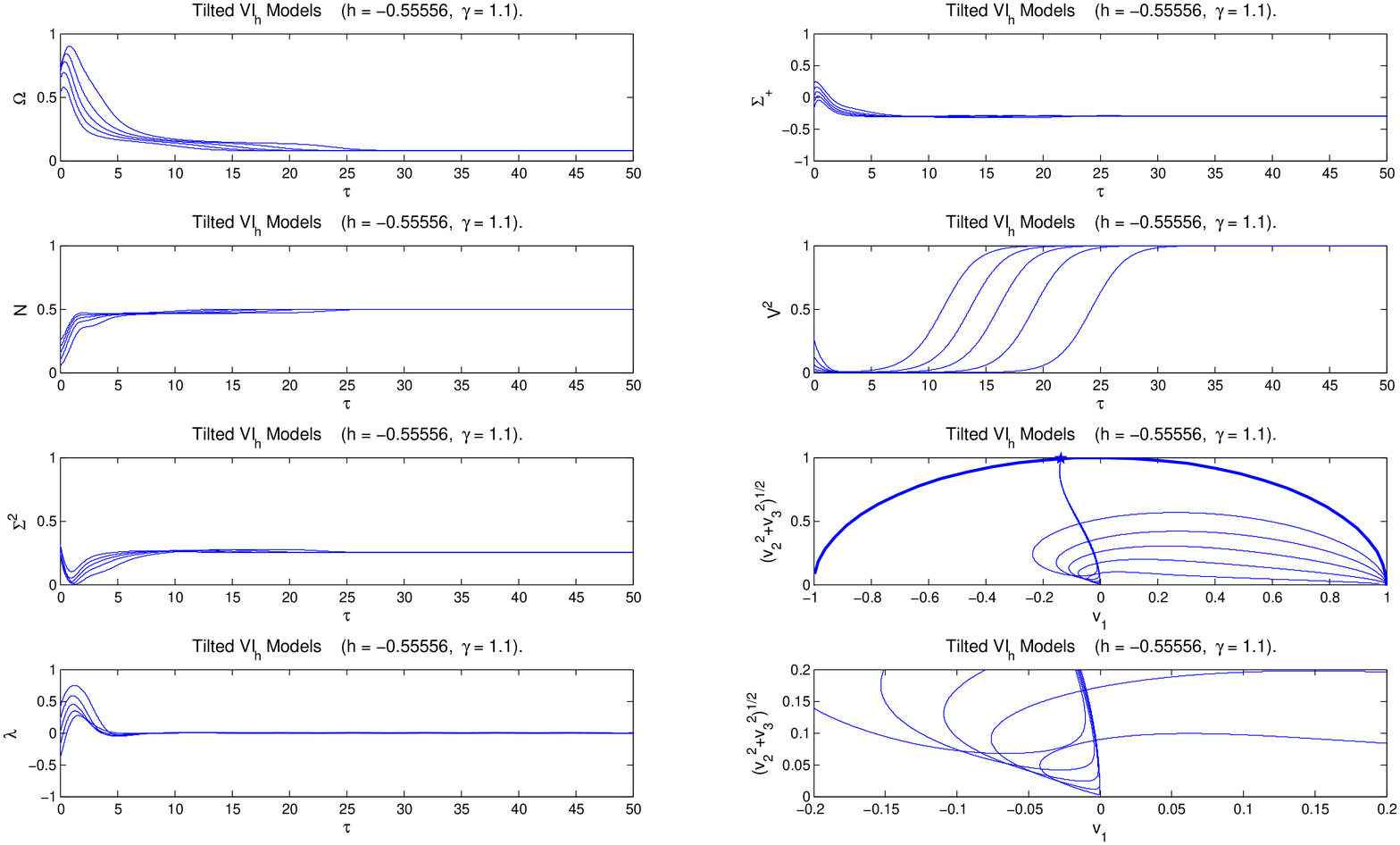}$$
\end{figure}

\begin{figure}[f]
\caption{Fully tilted Bianchi VI$_{h}$: $h=-4, \gamma=1.17$:  Attractor is $\tilde {\mathcal F}^{-}(h)$. The initial conditions are chosen to be $\Sigma_{+}= 0.3-\frac{i}{10},\Sigma_{-}= -0.3+\frac{i}{10},\Sigma_{12}=0.4,\Sigma_{13}=0.2,\Sigma_{23}=0.2,N= \frac{i}{40},\lambda=0.6-\frac{i}{5}$ with $i$ ranging from 1 to 4.}\label{FE2}
$$\includegraphics[scale=0.5,angle=90]{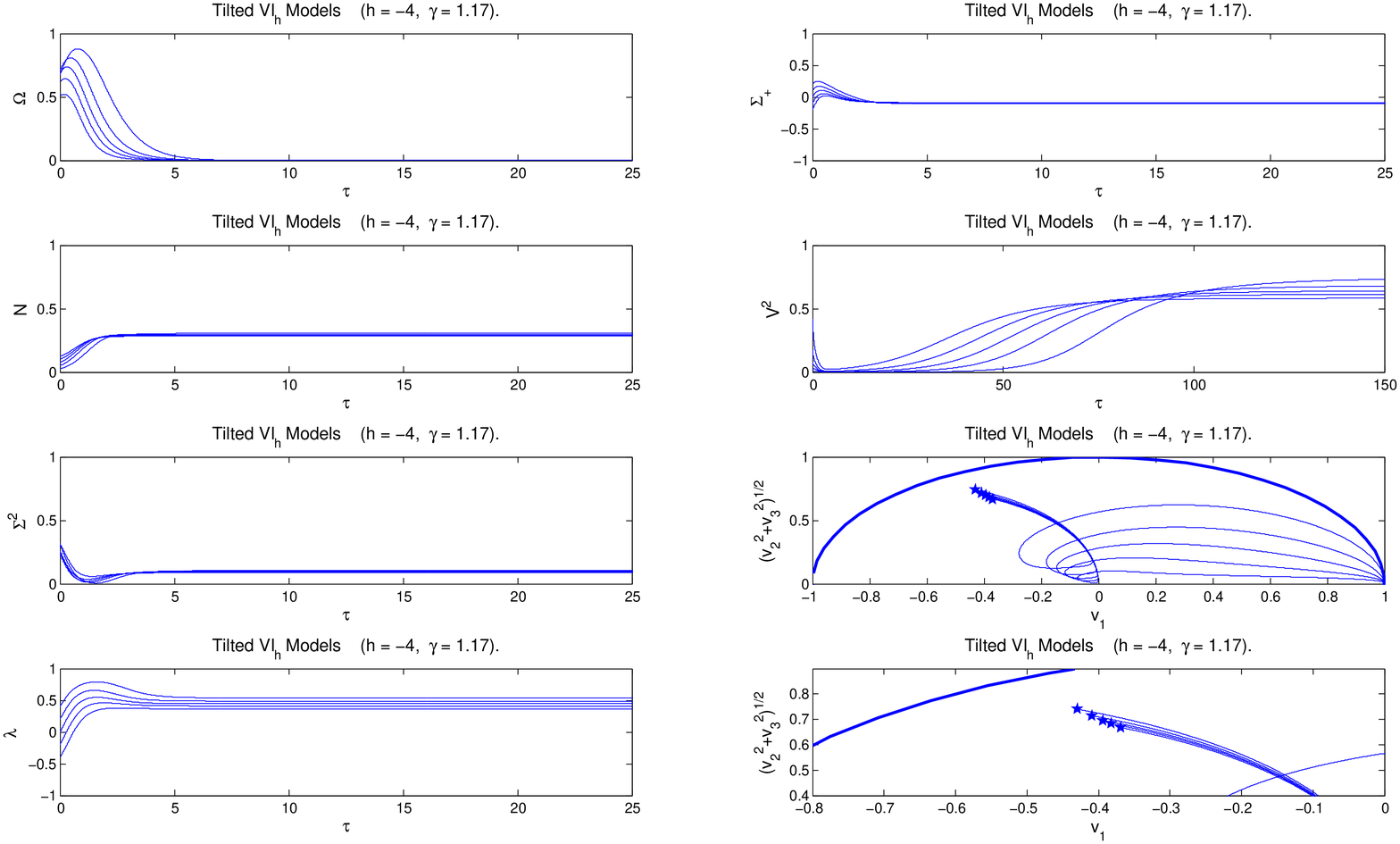}$$
\end{figure}

\begin{figure}[f]
\caption{Fully tilted Bianchi VI$_{h}$: $h=-4, \gamma=1.20$:  Attractor is $\tilde{\mathcal E}_p^-(h)$.The initial conditions are chosen to be $\Sigma_{+}= 0.3-\frac{i}{10},\Sigma_{-}= -0.3+\frac{i}{10},\Sigma_{12}=0.4,\Sigma_{13}=0.2,\Sigma_{23}=0.2,N= \frac{i}{40},\lambda=0.6-\frac{i}{5}$ with $i$ ranging from 1 to 5. }\label{FE3}
$$\includegraphics[scale=0.5,angle=90]{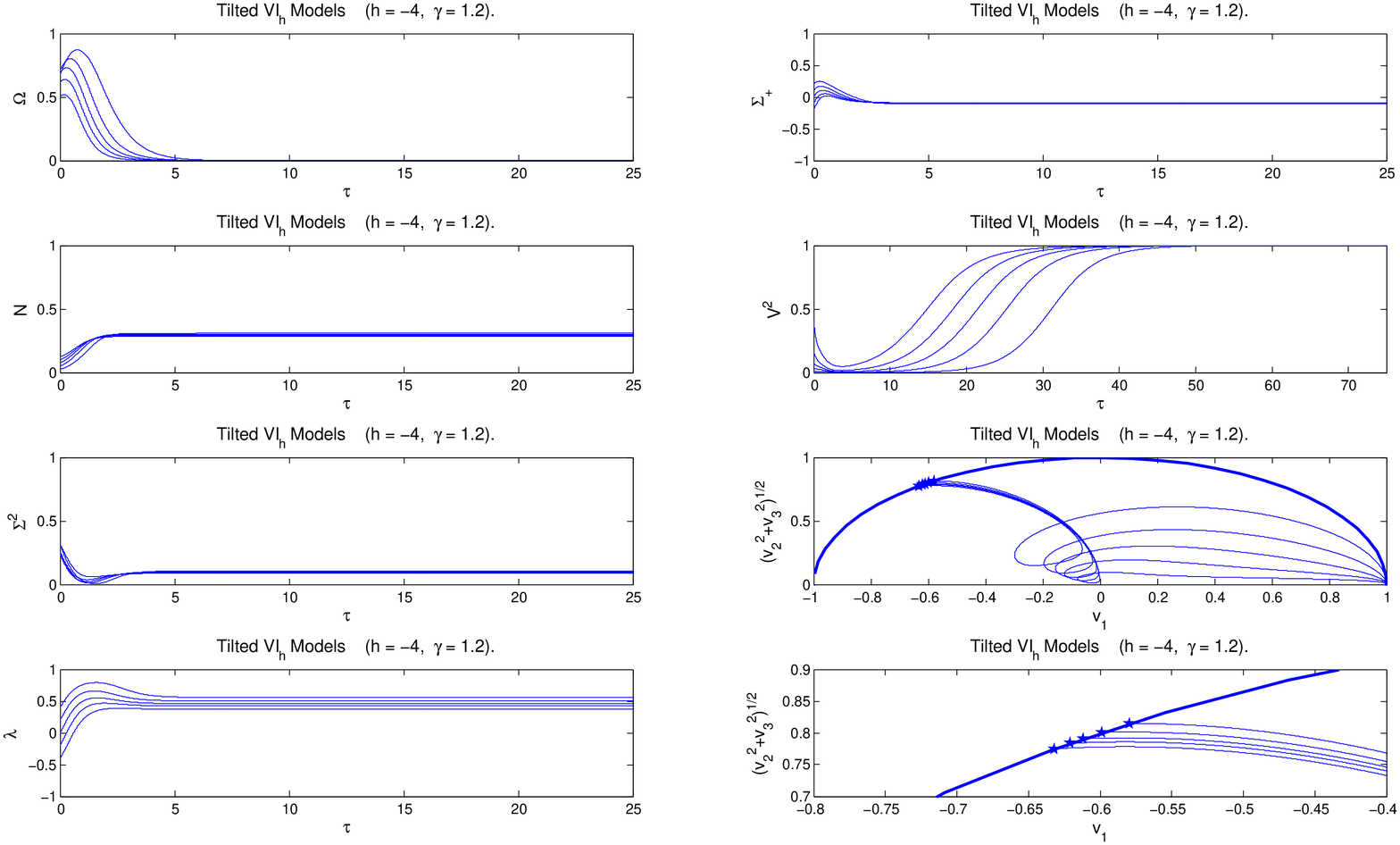}$$
\end{figure}

\begin{figure}[f]
\caption{Fully tilted Bianchi VI$_{h}$: $h=-9, \gamma=1.241$:  Attractor is $\tilde{\mathcal L}_{-}(h)$ and a closed orbit. The initial conditions are chosen to be $\Sigma_{+}= 0.3-\frac{i}{10},\Sigma_{-}= -0.3+\frac{i}{10},\Sigma_{12}=0.4,\Sigma_{13}=0.2,\Sigma_{23}=0.2,N= \frac{i}{60},\lambda=0.6-\frac{i}{5}$ with $i$ ranging from 1 to 5.}\label{FF6}
$$\includegraphics[scale=0.5,angle=90]{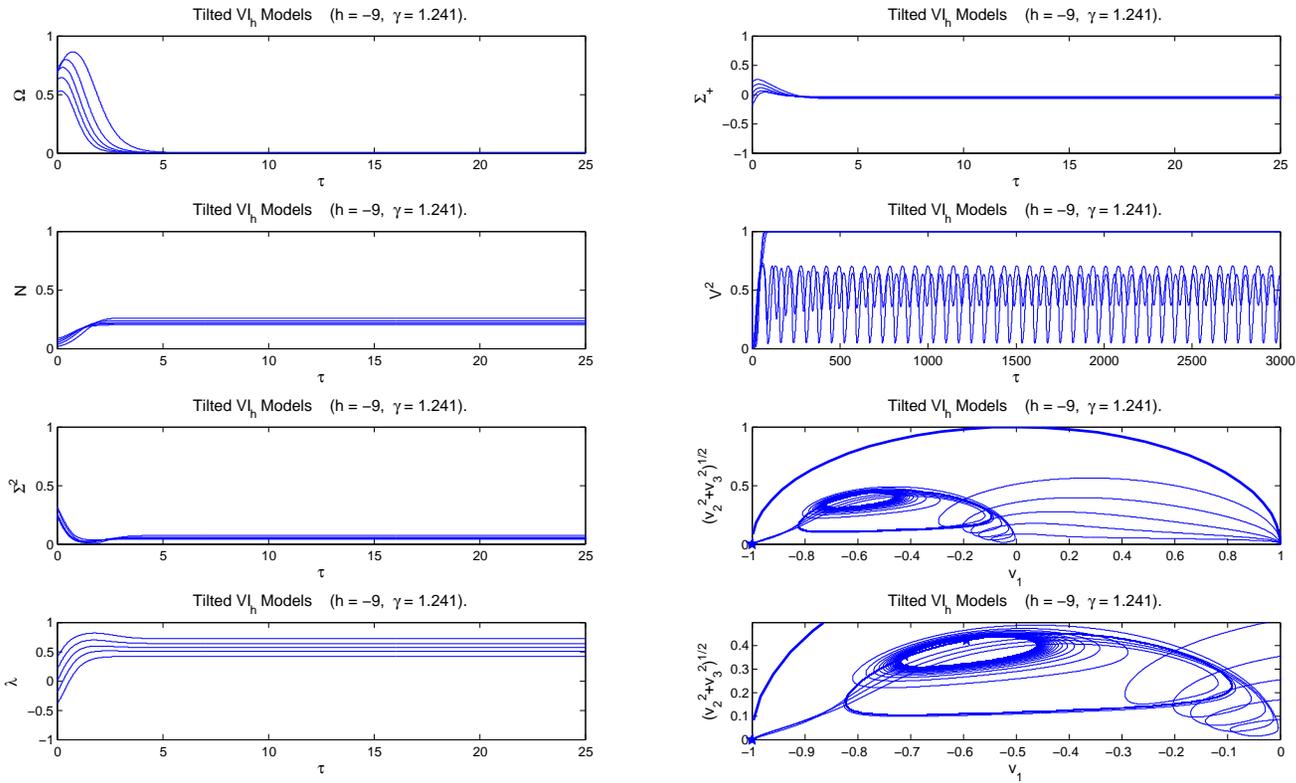}$$
\end{figure}

Typical values of the parameters $h$ and $\gamma$ are chosen to
illustrate some of the interesting future asymptotic behaviour found
in the fully tilted Bianchi VI$_h$ models, see Figures
\ref{FA4}-\ref{FF6}. For the interested reader, additional figures
can be found in the companion article \cite{companion2007}.  In each
numerical run we choose essentially the same set of initial
conditions unless otherwise indicated.  Given initial values for
$\{\Sigma_+,\Sigma_-,\Sigma_{12},\Sigma_{13},\Sigma_{23},N,\lambda\}$,
initial values for $A, v_1,v_2,v_3$ were determined by solving the
constraint equations.  For all cases except $h=-5/9$, the built in
differential equations solver {\em ode45} in Matlab R2006a was used
to numerically solve all cases.  The {\em Relative Tolerance} was set
to $10^{-12}$ and the {\em Absolute Tolerance} was set to
$10^{-20}$.  The full 11-dimensional set of differential equations
was integrated while the constraint equations were graphed (not
included here) and used to determine when and if the numerical
analysis breaks down.  In the case $h=-5/9$, due to instabilities of
the constraint surfaces, an alternative approach had to be taken.
The built in differential equations solver {\em ode15s} in Matlab
R2006a was used to numerically solve this case.  Instead of
integrating the 11-dimensional set of differential equations for the
quantities $\{\Sigma_+, \Sigma_-, \Sigma_{12}, \Sigma_{13},
\Sigma_{23}, N, \lambda, A, v_1, v_2, v_3\}$, a 7-dimensional set of
differential equations for $\{\Sigma_+, \Sigma_-, \Sigma_{12},
\Sigma_{13}, \Sigma_{23}, N, \lambda\}$ and a set of algebraic
constraints for $A,v_1,v_2,v_3$ were integrated. Again the
constraint equations were also graphed (not included here) and used
to determine when and if the numerical analysis break down.  The {\em
Relative Tolerance} was set to $10^{-12}$ and the {\em Absolute
Tolerance} was set to $10^{-15}$.    In each phase portrait stars
indicate the future attractor.

Figure \ref{FA4} shows a partial picture of the non-isolated nature
of the stable attractor $\mathcal{B}(h)$. In the $(v_1,\sqrt{v_2^2+v_3^2})$
phase portrait, the lower bound (as seen in the phase portrait) of
this non-isolated attractor is the equilibrium point $\mathcal{R}^-(h)$
(which is stable if the value of $\gamma$ is a little less than the
critical value of $6/5$) while the upper bound (not seen in the phase
portrait) of this non-isolated attractor is the equilibrium point
$\mathcal{E}_2(h)$ (which is stable if the value of $\gamma$ is larger
than the critical value of $6/5$).  Figure \ref{FC2} gives an
illustration of a situation in which there are two stable equilibrium
points in the fully tilted Bianchi VI$_h$ models.  Figure \ref{FC3}
shows the strongly attracting nature of the heteroclinic orbit
between $\mathcal{C}(h)$ and $\mathcal{E}_3(h)$.  Figure \ref{FE2} not only
shows the strongly attracting nature of the heteroclinic orbit
between $\mathcal{C}(h)$ and $\widetilde{\mathcal{F}}^-(h)$, but also the
non-isolated nature of the equilibrium point $\widetilde{\mathcal{F}}^-(h)$.
We also note how the variables $\{\Omega, N, \Sigma^2, \lambda\}$
obtain their equilibrium values much much faster than the tilt
velocities.  This behaviour is akin to the {\em freezing in}
behaviour observed in the type VII$_h$ and IV models
\cite{CH2,HHC}.  Again we observe this same {\em freezing in}
behaviour and the non-isolated nature of the equilibrium point
$\widetilde{\mathcal{E}}_p^-(h)$ in Figure \ref{FE3}. Figure \ref{FF6}
illustrates something that was not predicted in the previous
qualitative analysis; that is, the existence of two stable
attractors, a stable equilibrium point $\tilde{\mathcal{L}}_-(h)$ and a
stable closed orbit. The parameter values $h=-9, \gamma=1.241$ used
in Figure \ref{FF6} would be considered outside the 3-dimensional
analogue of the loophole.

\begin{table}
\centering
\begin{tabular}{|c|c|c|l|}
\hline
Bianchi &   & Future &  \\
Type & Matter & Attractor & Comments \\
\hline \hline
I & $\frac 23<\gamma<2$ & $\mathcal{I}(I)$ & No tilt allowed  \\
\hline
II & $\frac 23<\gamma<\frac {10}7$ & $\mathcal{C}\mathcal{S}(II)$ & Non-tilted Collins-Stewart \\
 & $\frac{10}7<\gamma<\frac{14}9$  & $\mathcal{H}(II)$ & Hewitt's tilted type II  \\
 & $\gamma=\frac{14}9$  & $\mathcal{L}(II)$ & Tilted type II bifurcation \\
 & $\frac{14}9<\gamma<2$  & $\mathcal{E}(II)$ & Extremely tilted \\
\hline
III & $\frac 23<\gamma<1$ & $\mathcal{C}(III)$ & Non-tilted Collins type VI$_{-1}$\\
 & $\gamma=1$ & $\mathcal{P}_0(III)$ & Non-tilted\\
 & $1<\gamma<2$ & $\mathcal{E}^-_p(III)$ & Extremely tilted\\ 
 \hline
IV & $\frac 23<\gamma<2$ & Plane waves & Tilted/non-tilted  \\
\hline
V & $\frac 23<\gamma<2$ & Milne &  Tilted/non-tilted  \\
\hline
$\underset{h<-1}{\text{VI}_{h}}$ & $\frac 23<\gamma<\frac{2(1-h)}{1-3h}$ & $\mathcal{C}(h)$ & Non-tilted Collins type VI$_h$ \\
  & $\frac{2(1-h)}{1-3h}<\gamma<2$ & Plane waves & Tilted/non-tilted  \\
\hline
$\underset{-1<h<-1/9}{\text{VI}_{h}}$ & $\frac 23<\gamma<1$& $\mathcal{C}(h)$ & Non-tilted Collins type VI$_h$ \\

&$ 1<\gamma<\frac{2(3+\sqrt{-h})}{5+3\sqrt{-h}}$ & $\mathcal{C}(h)~\& ~\mathcal{E}_3(h)$  & Non-tilted or extremely tilted  \\
  & $\frac{2(3+\sqrt{-h})}{5+3\sqrt{-h}}<\gamma<2$ &  $\mathcal{E}_3(h)$& Extremely tilted, non-vacuum.
\\
\hline
VI$_{-1/9}$ & $\frac 23<\gamma<\frac {10}9$ & $\mathcal{C}(h)$ & Non-tilted Collins type VI$_{-1/9}$ \\
 & $\gamma=\frac{10}9$ & $\mathcal{W}$ & Non-tilted Wainwright type VI$_{-1/9}$ \\
 & $\frac{10}9<\gamma< 2$ & $\widetilde{\mathcal{CF}}$ & Collinson-French: Tilted/non-tilted \\
\hline
$\underset{-1/9<h\leq 0}{\text{VI}_{h}}$ & $\frac 23<\gamma<\frac{2(3+\sqrt{-h})}{5+3\sqrt{-h}}$ & $\mathcal{C}(h)$ & Non-tilted Collins type VI$_h$ \\
& $\frac{2(3+\sqrt{-h})}{5+3\sqrt{-h}}<\gamma<\frac 65$ & $\mathcal{R}^-(h)$ & Intermediately tilted \\
& $ \gamma=\frac 65$ & $\mathcal{B}(h)$ & Tilted type VI$_h$, $\gamma=6/5$ bifurcation \\
& $\frac 65<\gamma<2$ & $\mathcal{E}_2(h)$ & Extremely tilted type VI$_h$\\
\hline
VII$_h$ & $\frac 23<\gamma<2$ & Plane waves & Tilted/non-tilted  \\
\hline
VII$_0$ & $\frac 23<\gamma<\frac 43$ & $\widetilde{P}_1(VII_0)$ & Non-tilted\\
 & $\gamma=\frac 43$ & $\widetilde{P}_3(VII_0)$ & Radiation bifurcation\\
 & $\frac 43<\gamma<2$ & $\widetilde{P}_4(VII_0)$ & Extremely tilted\\
\hline
VIII & $\frac 23<\gamma<1$ & $\widetilde{P}_1(VIII)$ & Non-tilted\\
 & $\gamma=1$ & $\widetilde{P}_2(VIII)$ & Non-tilted\\
 & $1<\gamma<2$ & $\widetilde{E}_1(VIII)$ & Extremely tilted\\ 
 \hline
\end{tabular}
\caption{The late-time behaviour of Bianchi  models with a tilted
$\gamma$-law perfect fluid (see the text for details and
references). The case $0<\gamma<2/3$ is covered by Corollary \ref{no-hair}.} \label{tab:outline}
\end{table}

\section{Discussion}

In this paper we have used dynamical systems methods and detailed
numerical experimentation to investigate the future asymptotic
properties of SH Bianchi type VI$_h$ cosmological models. We have
determined {\em all of the equilibrium points} of the type VI$_h$
state space. These equilibrium points correspond to exact
self-similar solutions of the Einstein equations (some of which
are {\em new} exact solutions) and play an important role in
describing the general evolution of the system. The stability of
all of these equilibrium points is also investigated. In
particular, we have determined all possible late time behaviours
for Bianchi type VI$_h$ models. All of the possible future
asymptotic behaviours for all Bianchi models (except Bianchi type
IX models, which can recollapse) are now known and summarized in
Table \ref{tab:outline}; this Table is now complete. We note that the Bianchi type
VI$_h$  case is of special interest in that it is very complicated
and contains many subcases, and many of the different future
asymptotic behaviours found in previous work occur in these
particular models.

In particular, it was found that the vacuum plane-wave solutions
play an important role in the future evolution of type VI$_h$
models (as was the case in the type VII$_h$ and the type IV
universes \cite{HHC}). All of the plane-wave equilibrium points
are found to occur in the invariant subspaces $N^{\pm}(VI_h)$. The
regions where the various plane-wave equilibrium points were
stable are depicted in Fig. \ref{FigT2}; we recall that for the
fully tilted models only the plane-waves in $N^-_2(VI_h)$ are
stable.

A loophole exists in which there are no stable equilibrium points,
and hence it is necessary to seek other candidates for late-time
attractors. It was noted that closed curves and tori act as
attractors in the corresponding loopholes in type IV and type
VII$_h$ models \cite{CH2,HHC}. Hence, we looked for closed curves
in the type VI$_h$ loophole. Calculations and numerical
experimentation were found to provide plausible evidence that the
system experiences a Hopf-bifurcation resulting in a stable closed
orbit as $\gamma$ takes values in a range of parameter values
(that includes values within the loophole, but also includes
values just outside the loophole). This shows that there are
attracting closed curves even outside the loophole; however,
outside the loophole these attracting curves will co-exist with
attracting equilibrium points and there are consequently a number
of possible late time behaviours. Inside the loophole, however,
only the closed curves can be attactors since all of the
equilibrium points in the loophole are unstable. The family of
attracting closed orbits (both outside and inside the loophole),
are referred to as the Mussel attractor.

In more detail, we analytically proved that in the type VI$_h$
loophole there exists a closed periodic orbit (the Mussel
attractor), $\mathcal{C}_M(h)$, for the dynamical system. In
addition, we provided convincing numerical evidence that closed
orbits exist outside of loophole (see Figs. \ref{T2_fig_1}
and \ref{T2_fig_2}). Clearly, an analysis of the equilibrium points
alone is not sufficient for determining the future asymptotic
behaviour in the Bianchi type VI$_h$ models. However, we should
also note that apart from the the Mussel attractor we have found
no evidence for any other closed period orbits important for the
late-time behaviour for $h\neq -1/9$.

The Bianchi type III case (i.e., $h=-1$) is a special case which
necessitates a separate treatment. For $2/3<\gamma<1$ the
Collins solution is the only attractor. However, for $\gamma \ge
1$, the corresponding equilibrium points have multiple
zero-eigenvalues, and a centre manifold analysis is required.
We found that, in the invariant subspace $T^-_2(VI_h)$ for $1<\gamma<2$, the type III models approach the
extremely tilted vacuum solution
$\widetilde{\mathcal{E}}_p^-(III)$  at late times. In addition,
based upon a detailed centre manifold analysis in the invariant
subspace $T^-_2(VI_h)$ and numerical simulations in the fully tilted
space $T(VI_h)$, we concluded that it is plausible that
$\mathcal{P}_0(III)$ is the attractor for the fully tilted dust
($\gamma=1$) type III model. \footnote{We should also point out that the analysis of the special Bianchi type VI$_{-1/9}$ case is not complete either. Due to the exact vanishing of one of the contraint equations a different approach has to be taken regarding the numerical analysis.  Furthermore, the question whether there are closed periodic orbits acting as attractors in this case has not been addressed in this paper; however, a prelimiary analysis indicates that such attracting closed curves do indeed exist \cite{HHLC19}.}

As noted above, comprehensive numerical experiments were carried
out to complement and confirm the analytical results. All of the
numerics, which serve as confirmation of all of our analytical
results, are summarized in a companion article \cite{companion2007}. In the
present paper we have included several figures (from \cite{companion2007}), that
provide typical results (confirming the analytical results) and
some specific figures that contain interesting phenomena.


\section*{Acknowledgements} 
This work was supported by an AARMS Postdoctoral Fellowship (SH), CIAR (WCL) and 
the Natural Sciences and Engineering Research Council of Canada (SH, RJvdH, WCL and AAC). 
\appendix

\bibliographystyle{amsplain}
\appendix
\section{Nasty expressions}
\subsection{Expressions determining $V$, $\phi$ and $\lambda$ for $\mathcal{B}(h)$} 
The following expressions are produced using \textsc{Maple}. 
\begin{multline}
V^2=2\,N^{2}( - 1728 - 724275\,N^{6}\,r^{6}\,k^{4} + 178848\,N^{4}\,k
^{4}\,r^{4} - 11664\,k^{4}\,r^{4}\,N^{2} - 900\,N^{6}\,r^{8}\,k^{
2} \\
\mbox{} + 1350\,N^{8}\,r^{10}\,k^{4} + 26244\,N^{6}\,k^{8}\,r^{8}
 + 32940\,k^{4}\,r^{8}\,N^{6} - 39690\,k^{6}\,r^{10}\,N^{8} \\
\mbox{} + 13122\,k^{10}\,r^{10}\,N^{8} - 625\,N^{6}\,r^{6} +
34560\,N^{2} - 31104\,k^{2}\,r^{2}\,N^{2} + 150\,N^{4}\,r^{6} \\
\mbox{} + 80000\,N^{6} + 368550\,N^{8}\,k^{6}\,r^{8} - 525600\,k
^{2}\,r^{4}\,N^{6} - 170640\,N^{4}\,k^{2}\,r^{2} \\
\mbox{} - 25000\,k^{2}\,r^{6}\,N^{8} + 414720\,N^{6}\,k^{4}\,r^{4
} + 72075\,N^{6}\,k^{2}\,r^{6} - 77031\,N^{6}\,k^{6}\,r^{6} -
100800\,N^{4} \\
\mbox{} + 2592\,k^{2}\,r^{2} + 864\,r^{2} - 90000\,N^{6}\,r^{2}
 - 126846\,N^{8}\,k^{8}\,r^{10} - 8730\,N^{4}\,k^{2}\,r^{6} \\
\mbox{} + 320400\,k^{4}\,r^{6}\,N^{8} - 27054\,N^{4}\,r^{6}\,k^{4
} + 1458\,N^{4}\,k^{6}\,r^{6} - 1728\,N^{2}\,r^{4}\,k^{2} \\
\mbox{} - 259200\,k^{4}\,r^{4}\,N^{8} + 80000\,k^{2}\,r^{4}\,N^{8
} + 358800\,k^{2}\,r^{2}\,N^{6} + 270000\,k^{2}\,r^{4}\,N^{4} \\
\mbox{} - 28800\,r^{2}\,N^{2} - 13200\,N^{4}\,r^{4} + 20000\,N^{6
}\,r^{4} + 99600\,N^{4}\,r^{2} + 1250\,N^{8}\,r^{8}\,k^{2} \\
\mbox{} + 56862\,N^{8}\,k^{8}\,r^{8} + 115668\,N^{6}\,r^{8}\,k^{6
} - 236520\,N^{8}\,k^{6}\,r^{6} - 80550\,k^{4}\,r^{8}\,N^{8} +
720\,N^{2}\,r^{4} \\
)/[(9\,k^{2}\,r^{2}\,N^{2} - 3 + 5\,N^{2})(5103\,N^{6}\,k^{6}\,r
^{6} - 50625\,N^{6}\,r^{6}\,k^{4} + 40500\,N^{6}\,k^{4}\,r^{4}
 \\
\mbox{} + 10692\,N^{4}\,k^{4}\,r^{4} - 2475\,N^{6}\,k^{2}\,r^{6}
 - 66600\,k^{2}\,r^{4}\,N^{6} + 37800\,k^{2}\,r^{4}\,N^{4} \\
\mbox{} + 50400\,k^{2}\,r^{2}\,N^{6} - 32400\,N^{4}\,k^{2}\,r^{2}
 + 1296\,k^{2}\,r^{2}\,N^{2} + 125\,N^{6}\,r^{6} + 500\,N^{6}\,r
^{4} \\
\mbox{} - 300\,N^{4}\,r^{4} - 12000\,N^{6}\,r^{2} + 13200\,N^{4}
\,r^{2} - 3600\,r^{2}\,N^{2} + 32000\,N^{6} - 52800\,N^{4} \\
\mbox{} + 28800\,N^{2} - 5184)] 
\label{Vsq}
\end{multline}
\begin{multline}
\cos(2\phi)=
3(13851\,N^{6}\,k^{6}\,r^{6} - 144585\,N^{6}
\,r^{6}\,k^{4} - 10575\,N^{6}\,k^{2}\,r^{6} + 125\,N^{6}\,r^{6}
 \\
\mbox{} + 116640\,N^{6}\,k^{4}\,r^{4} - 201600\,N^{6}\,r^{4}\,k^{
2} - 4000\,N^{6}\,r^{4} + 176400\,k^{2}\,r^{2}\,N^{6} - 58000\,N
^{6}\,r^{2} \\
\mbox{} + 80000\,N^{6} + 11664\,N^{4}\,k^{4}\,r^{4} + 127440\,k^{
2}\,r^{4}\,N^{4} + 2400\,N^{4}\,r^{4} - 88560\,k^{2}\,r^{2}\,N^{4
} \\
\mbox{} + 73200\,N^{4}\,r^{2} - 100800\,N^{4} - 10368\,k^{2}\,r^{
2}\,N^{2} - 23040\,N^{2}\,r^{2} + 34560\,N^{2} - 1728)r \\
k/( - 1728 + 34560\,N^{2} - 326160\,k^{2}\,r^{2}\,N^{4} + 864\,r
^{2} - 7776\,k^{2}\,r^{2} + 41472\,k^{2}\,r^{2}\,N^{2} \\
\mbox{} - 90000\,N^{6}\,r^{2} - 13200\,N^{4}\,r^{4} - 8370\,N^{4}
\,k^{2}\,r^{6} + 99600\,N^{4}\,r^{2} - 100800\,N^{4} \\
\mbox{} - 28800\,N^{2}\,r^{2} + 464400\,k^{2}\,r^{2}\,N^{6} +
20000\,N^{6}\,r^{4} - 625\,N^{6}\,r^{6} + 11664\,k^{4}\,r^{4}\,N
^{2} \\
\mbox{} - 7776\,N^{4}\,k^{4}\,r^{4} + 30618\,N^{4}\,k^{6}\,r^{6}
 + 80000\,N^{6} + 505440\,N^{6}\,k^{4}\,r^{4} + 152361\,N^{6}\,k
^{6}\,r^{6} \\
\mbox{} + 56475\,N^{6}\,k^{2}\,r^{6} - 702675\,N^{6}\,r^{6}\,k^{4
} + 59778\,N^{4}\,r^{6}\,k^{4} - 7776\,N^{2}\,r^{4}\,k^{2} \\
\mbox{} + 270000\,k^{2}\,r^{4}\,N^{4} - 460800\,N^{6}\,r^{4}\,k^{
2} - 132678\,N^{6}\,r^{8}\,k^{6} + 13122\,N^{6}\,k^{8}\,r^{8} \\
\mbox{} - 450\,N^{6}\,r^{8}\,k^{2} + 18630\,N^{6}\,r^{8}\,k^{4}
 + 720\,N^{2}\,r^{4} + 150\,N^{4}\,r^{6})
 \label{cos2phiC2}
 \end{multline}
Polynomial:
\begin{multline}
P(k,N,\lambda)\equiv- 9( - 24 + 130\,N^{2} - 125\,N^{4} + 25\,N^{
4}\,k + 135\,N^{4}\,k^{3} + 54\,N^{4}\,k^{4} + 54\,k^{2}\,N^{2}
 \\
\mbox{} + 135\,N^{4}\,k^{2} - 120\,N^{2}\,k)( - 24 + 130\,N^{2}
 - 125\,N^{4} - 25\,N^{4}\,k - 135\,N^{4}\,k^{3} + 54\,N^{4}\,k^{
4} \\
\mbox{} + 54\,k^{2}\,N^{2} + 135\,N^{4}\,k^{2} + 120\,N^{2}\,k)(
12 + 9\,k^{2}\,N^{2} - 25\,N^{2})^{2} + 3\,N^{2}(414720 \\
\mbox{} + 26611200\,N^{2} + 229770000\,N^{6} - 132408000\,N^{4}
 - 746496\,k^{2} \\
\mbox{} - 403734375\,N^{10}\,k^{2} - 47202750\,k^{6}\,N^{10} +
6298560\,N^{4}\,k^{6} + 40234375\,N^{10} \\
\mbox{} + 550314000\,k^{4}\,N^{8} - 165500000\,N^{8} - 38283435\,
N^{10}\,k^{10} + 93658275\,N^{10}\,k^{8} \\
\mbox{} - 341388000\,k^{2}\,N^{6} - 28868400\,N^{8}\,k^{8} -
331840800\,N^{8}\,k^{6} + 115998750\,N^{10}\,k^{4} \\
\mbox{} + 725805000\,N^{8}\,k^{2} + 59486400\,N^{4}\,k^{4} -
7776000\,k^{2}\,N^{2} + 39528000\,N^{4}\,k^{2} \\
\mbox{} - 229780800\,N^{6}\,k^{4} + 78732000\,N^{6}\,k^{6} -
2799360\,N^{2}\,k^{4} + 16533720\,N^{8}\,k^{10} \\
\mbox{} + 4251528\,N^{10}\,k^{12} + 20470320\,N^{6}\,k^{8})
{\lambda}^{2} - 10\,N^{4}(77760 + 3888000\,N^{2} \\
\mbox{} + 10462500\,N^{6} - 12352500\,N^{4} - 1399680\,k^{2} +
9211644\,N^{4}\,k^{8} + 25194240\,N^{4}\,k^{6} \\
\mbox{} - 56457000\,k^{4}\,N^{8} - 419904\,k^{4} - 1484375\,N^{8}
 + 339174000\,k^{2}\,N^{6} \\
\mbox{} + 104090265\,N^{8}\,k^{8} - 34736850\,N^{8}\,k^{6} -
206015625\,N^{8}\,k^{2} + 1889568\,N^{2}\,k^{6} \\
\mbox{} - 61148520\,N^{4}\,k^{4} - 2060640\,k^{2}\,N^{2} -
130442400\,N^{4}\,k^{2} + 414752400\,N^{6}\,k^{4} \\
\mbox{} - 211789080\,N^{6}\,k^{6} + 14696640\,N^{2}\,k^{4} +
3188646\,N^{8}\,k^{12} - 35724645\,N^{8}\,k^{10} \\
\mbox{} - 32673780\,N^{6}\,k^{8} + 9920232\,N^{6}\,k^{10})
{\lambda}^{4} + 30\,N^{6}(7200 + 64044000\,N^{4}\,k^{2} \\
\mbox{} - 19306350\,N^{6}\,k^{6} - 18994095\,N^{6}\,k^{10} -
72171000\,N^{4}\,k^{6} + 1400000\,N^{4} \\
\mbox{} - 2799360\,N^{2}\,k^{4} + 64921095\,N^{6}\,k^{8} -
50521875\,k^{2}\,N^{6} - 64739250\,N^{6}\,k^{4} \\
\mbox{} - 339000\,N^{2} + 3306744\,N^{4}\,k^{10} - 16008840\,N^{4
}\,k^{8} + 1417176\,N^{6}\,k^{12} - 375840\,k^{2} \\
\mbox{} + 2047032\,N^{2}\,k^{8} + 3324240\,N^{2}\,k^{6} +
132013800\,N^{4}\,k^{4} - 22366800\,k^{2}\,N^{2} \\
\mbox{} + 1283040\,k^{4} + 209952\,k^{6} - 1328125\,N^{6})
{\lambda}^{6} - 15\,N^{8}(13599000\,k^{2}\,N^{2} \\
\mbox{} - 33934950\,N^{4}\,k^{6} - 33165855\,N^{4}\,k^{10} -
71092080\,N^{2}\,k^{6} - 262500\,N^{2} \\
\mbox{} + 4694760\,k^{4} + 130793535\,N^{4}\,k^{8} - 27496875\,N
^{4}\,k^{2} - 99636750\,N^{4}\,k^{4} + 1500 \\
\mbox{} + 3306744\,N^{2}\,k^{10} - 21126420\,N^{2}\,k^{8} +
2125764\,N^{4}\,k^{12} + 1023516\,k^{8} + 524880\,k^{6} \\
\mbox{} + 88176600\,N^{2}\,k^{4} - 183600\,k^{2} + 546875\,N^{4})
{\lambda}^{8} + N^{10}\,(9\,k^{2} - 30\,k + 5) \\
(9\,k^{2} + 30\,k + 5)(157464\,N^{2}\,k^{8} + 122472\,k^{6} -
1228365\,N^{2}\,k^{6} + 232875\,N^{2}\,k^{4} \\
\mbox{} + 252720\,k^{4} - 5400\,k^{2} + 410625\,k^{2}\,N^{2} +
15625\,N^{2}){\lambda}^{10} -  \\
324\,N^{12}\,k^{4}\,(9\,k^{2} - 30\,k + 5)^{2}\,(9\,k^{2} + 30\,k
 + 5)^{2}\,{\lambda}^{12} 
\label{Polynomial}\end{multline}
\subsection{Expressions determining $\Omega$ and $V$ for $\mathcal{R}(h)$} 
\begin{multline}
\Omega = - (24 - 92\,\gamma  + 120\,k + 168\,k^{2} + 72\,k^{3} +
114\,\gamma ^{2} - 45\,\gamma ^{3} + 96\,{\Sigma_+}^{2} - 96\,
{\Sigma_+} - 376\,k\,\gamma  \\
\mbox{} + 366\,k^{2}\,\gamma ^{2} + 366\,k\,\gamma ^{2} - 444\,k
^{2}\,\gamma  - 93\,k^{2}\,\gamma ^{3} - 108\,k\,\gamma ^{3} - 90
\,k^{3}\,\gamma ^{3} + 258\,k^{3}\,\gamma ^{2} \\
\mbox{} - 240\,k^{3}\,\gamma  - 672\,k^{2}\,{\Sigma_+} - 480\,k\,
{\Sigma_+} + 480\,k\,{\Sigma_+}^{2} + 672\,k^{2}\,{\Sigma_+}^{2} - 288
\,k^{3}\,{\Sigma_+} \\
\mbox{} + 288\,k^{3}\,{\Sigma_+}^{2} + 224\,\gamma \,{\Sigma_+} - 80
\,{\Sigma_+}^{2}\,\gamma  - 120\,\gamma ^{2}\,{\Sigma_+} + 1632\,k^{2
}\,{\Sigma_+}\,\gamma  + 276\,{\Sigma_+}\,k^{2}\,\gamma ^{3} \\
\mbox{} + 1168\,k\,\gamma \,{\Sigma_+} + 1200\,{\Sigma_+}^{2}\,k^{2}
\,\gamma ^{2} - 1224\,{\Sigma_+}\,k^{2}\,\gamma ^{2} - 948\,k\,
\gamma ^{2}\,{\Sigma_+} - 832\,k\,\gamma \,{\Sigma_+}^{2} \\
\mbox{} - 1488\,k^{2}\,\gamma \,{\Sigma_+}^{2} - 360\,{\Sigma_+}^{2}
\,k^{2}\,\gamma ^{3} + 270\,k\,\gamma ^{3}\,{\Sigma_+} + 360\,k\,
\gamma ^{2}\,{\Sigma_+}^{2} + 126\,{\Sigma_+}\,k^{3}\,\gamma ^{3} \\
\mbox{} - 444\,{\Sigma_+}\,k^{3}\,\gamma ^{2} + 216\,{\Sigma_+}^{2}\,
k^{3}\,\gamma ^{3} - 216\,{\Sigma_+}^{2}\,k^{3}\,\gamma ^{2} + 624
\,k^{3}\,{\Sigma_+}\,\gamma  - 288\,k^{3}\,{\Sigma_+}^{2}\,\gamma )
 \left/ {\vrule height0.43em width0em depth0.43em} \right. \!
 \! \big{[}6 \\
\gamma ^{2}\,k^{2}\,(6 + 3\,k\,\gamma  - 5\,\gamma  - 2\,k)\big{]}
\label{OmegaC4}
\end{multline}
\begin{multline}
V^2=(24 - 92\,\gamma  + 144\,k + 216\,k^{2} + 114\,\gamma ^{2} - 45
\,\gamma ^{3} + 96\,{\Sigma_+}^{2} - 96\,{\Sigma_+} - 432\,k\,\gamma
 + 378\,k^{2}\,\gamma ^{2} \\
\mbox{} + 396\,k\,\gamma ^{2} - 492\,k^{2}\,\gamma  - 93\,k^{2}\,
\gamma ^{3} - 108\,k\,\gamma ^{3} - 90\,k^{3}\,\gamma ^{3} + 168
\,k^{3}\,\gamma ^{2} - 72\,k^{3}\,\gamma  \\
\mbox{} - 864\,k^{2}\,{\Sigma_+} - 576\,k\,{\Sigma_+} + 576\,k\,
{\Sigma_+}^{2} + 864\,k^{2}\,{\Sigma_+}^{2} + 224\,\gamma \,{\Sigma_+}
 - 80\,{\Sigma_+}^{2}\,\gamma  \\
\mbox{} - 120\,\gamma ^{2}\,{\Sigma_+} + 1968\,k^{2}\,{\Sigma_+}\,
\gamma  + 276\,{\Sigma_+}\,k^{2}\,\gamma ^{3} + 1320\,k\,\gamma \,
{\Sigma_+} + 1440\,{\Sigma_+}^{2}\,k^{2}\,\gamma ^{2} \\
\mbox{} - 1368\,{\Sigma_+}\,k^{2}\,\gamma ^{2} - 1008\,k\,\gamma ^{
2}\,{\Sigma_+} - 912\,k\,\gamma \,{\Sigma_+}^{2} - 1968\,k^{2}\,
\gamma \,{\Sigma_+}^{2} - 360\,{\Sigma_+}^{2}\,k^{2}\,\gamma ^{3} \\
\mbox{} + 270\,k\,\gamma ^{3}\,{\Sigma_+} + 360\,k\,\gamma ^{2}\,
{\Sigma_+}^{2} + 126\,{\Sigma_+}\,k^{3}\,\gamma ^{3} - 192\,{\Sigma_+}
\,k^{3}\,\gamma ^{2} + 216\,{\Sigma_+}^{2}\,k^{3}\,\gamma ^{3} \\
\mbox{} - 360\,{\Sigma_+}^{2}\,k^{3}\,\gamma ^{2} + 72\,k^{3}\,
{\Sigma_+}\,\gamma  + 144\,k^{3}\,{\Sigma_+}^{2}\,\gamma ) \left/
{\vrule height0.43em width0em depth0.43em} \right. \!  \! \big{[} \\
(6\,k^{2}\,{\Sigma_+}\,\gamma  + 3\,\gamma  - 2 + 4\,{\Sigma_+} + 5\,
k\,\gamma  - 10\,k\,\gamma \,{\Sigma_+} - 6\,k + 12\,k\,{\Sigma_+})(
60\,k\,\gamma ^{2}\,{\Sigma_+} \\
\mbox{} - 21\,k\,\gamma ^{2} - 15\,\gamma ^{2} + 60\,k\,\gamma
 + 28\,\gamma  - 132\,k\,\gamma \,{\Sigma_+} - 20\,\gamma \,\Sigma_+
 + 72\,k\,{\Sigma_+} - 36\,k - 12 \\
\mbox{} + 24\,{\Sigma_+})\big{]}
\label{VsqC4}
\end{multline}

\end{document}